\documentclass[12pt]{article}

\font\grande=cmr9.5 scaled \magstep4
\font\medio=cmr9.5 scaled \magstep2
\outer\def\beginsection#1\par{\medbreak\bigskip
      \message{#1}\leftline{\bf#1}\nobreak\medskip
\vskip-\parskip
      \noindent}
\usepackage{graphicx} 
\begin{document}
\bibliographystyle {unsrt}

\titlepage

\begin{flushright}
CERN-PH-TH/2012-055
\end{flushright}

\vspace{10mm}
\begin{center}
{\grande Fluid phonons and inflaton quanta}\\
\vspace{0.5cm}
{\grande at the protoinflationary transition}\\
\vspace{1.5cm}
 Massimo Giovannini
 \footnote{Electronic address: massimo.giovannini@cern.ch}\\
\vspace{1cm}
{{\sl Department of Physics, 
Theory Division, CERN, 1211 Geneva 23, Switzerland }}\\
\vspace{0.5cm}
{{\sl INFN, Section of Milan-Bicocca, 20126 Milan, Italy}}
\vspace*{0.5cm}
\end{center}

\vskip 0.5cm
\centerline{\medio  Abstract}
Quantum and thermal fluctuations of an irrotational fluid are studied across the transition regime connecting a
protoinflationary phase of decelerated expansion to an accelerated epoch driven by a single inflaton field. The protoinflationary inhomogeneities are suppressed when the transition to the slow roll phase occurs sharply over space-like hypersurfaces of constant energy density. If the transition is delayed, the interaction of the quasi-normal modes related, asymptotically, to fluid phonons and inflaton quanta leads to an enhancement of curvature perturbations.  It is shown that the dynamics of the fluctuations across the protoinflationary boundaries is determined by the monotonicity properties of the pump fields controlling the energy transfer between the background geometry and the quasi-normal modes of the fluctuations.  After corroborating the analytical arguments with explicit numerical examples, general lessons are drawn on the classification of the protoinflationary 
transition. 
\vskip 0.5cm

\noindent

\vspace{5mm}

\vfill
\newpage
\renewcommand{\theequation}{1.\arabic{equation}}
\setcounter{equation}{0}
\section{Introduction}
\label{sec1}
In the conventional lore, the large-scale temperature and polarization anisotropies of the Cosmic Microwave Background are caused by curvature inhomogeneities with typical wavelengths exceeding the Hubble radius at the time of matter radiation equality \cite{wmap1,wmap2}.  A nearly flat spectrum of Gaussian
fluctuations of the spatial curvature naturally arises from the quantum inhomogeneities of a single inflaton field evolving during a quasi-de Sitter stage of expansion. Although the simplest scenario is consistent with the observational signatures, different sets of initial conditions have been explored through the years.

Initial states different from the vacuum can modify the temperature and polarization anisotropies at large scales. This general approach has been scrutinized along various perspectives (see, e.g. \cite{one,two,three,four,five,six,seven,eight,nine}). Temperature-dependent phase transitions \cite{one,two} lead to an initial thermal state 
for the metric perturbations \cite{three,four,five,seven,eight}. If the initial state is not thermal (but it is 
not the vacuum either), curvature phonons can be similarly produced via stimulated emission. Second-order correlation effects of the scalar and tensor fluctuations of the geometry can be used to explore the statistical properties of the initial quantum state \cite{six} by applying the tenets of Hanbury-Brown-Twiss interferometry \cite{QO} which is employed, in quantum optics, to infer the bunching properties of visible light.

The modifications of the initial state are subjected to a number of constraints all originating, directly or indirectly, from the comparison between the energetic content of the initial fluctuations and the energy density of the background geometry. 
The criterion for the avoidance of severe backreaction effects is not unique. Single field quasi-de Sitter inflationary models with general initial states of primordial quantum fluctuations have been examined in \cite{seven,nine} with the aim of deriving constraints from the study of higher-order correlation functions and from the renormalizability of the energy-momentum tensor of the fluctuations. It is equally plausible to demand that the energy-momentum pseudo-tensor of the scalar and tensor fluctuations does not exceed the energy density and pressure of the background geometry, as argued in \cite{eleven}.

In the present paper a complementary and novel approach to the problem 
of the initial conditions of cosmological perturbations is pursued. The ever expanding inflationary backgrounds are geodesically 
incomplete and inflation cannot be eternal in the past. Thus it is legitimate to suppose the existence of a protoinflationary phase where the dynamics of the background was not yet accelerated. As the terminology indicates, the purpose here is not to test the universality of inflation given a set of arbitrary and widely different choices of the preinflationary dynamics. While under certain conditions inflation can be dynamically realized,  it is not eternal in the past either. The modest purpose here is not to challenge inflation but to analyze the initial conditions of large-scale inhomogeneities in an improved dynamical framework. Just to avoid potential misunderstandings, 
it should be clear that the transition from a deceleration to acceleration has nothing to do with the so-called bouncing behaviour where the background passes from contraction to expansion or vice versa (see, e.g. \cite{bounce} and references therein). In the present framework the universe will always be expanding even during the protoinflationary phase. 

The approach suggested here is pragmatic and the attention is focused on
single field inflationary models leading to a nearly flat spectrum of curvature inhomogeneities \cite{haw}.
During the  protoinflationary phase of decelerated dynamics the energy momentum tensor is dominated by a single perfect fluid. The analysis can be generalized to  include various inflaton fields and protoinflationary fluids but this 
will not be the primary goal of this investigation. 
Unlike the standard scenario, during the protoinflationary phase the seeds of curvature inhomogeneities are fluid phonons, i.e. the quantum excitations of an irrotational and relativistic fluid discussed by Lukash \cite{lukash} (see also \cite{lif,strokov}) right after one of the first formulations of inflationary dynamics \cite{staro1}. The whole irrotational system can be reduced 
to a single (decoupled) normal mode which is promoted to field operator in case the initial fluctuations are required to minimize the quantum Hamiltonian of the phonons \cite{lukash}. The canonical normal mode identified in \cite{lukash}  is invariant under infinitesimal coordinate transformations as required in the context of the  Bardeen formalism \cite{bard1} (see also \cite{lif}). The subsequent analyses of Refs.  \cite{KS} and  \cite{chibisov} follow the same logic of \cite{lukash} but in the case of scalar field matter;  the normal modes identified in \cite{lukash,KS,chibisov}  coincide with the (rescaled) curvature perturbations on comoving orthogonal hypersurfaces \cite{br1,bard2} (see also the beginning of section \ref{sec3}).  

The fluid phonons can be treated quantum mechanically but the initial state does not need to be the vacuum: if the protoinflationary phase is dominated by radiation, the fluid phonons are more likely to follow a Bose-Einstein distribution as it happens in inflationary models based on temperature-dependent phase transitions \cite{two,three,four}. When the protoinflationary  inhomogeneities are suppressed across the boundary, 
the initial normalization of curvature perturbations is set, most likely, by the quantum mechanical fluctuations generated during the inflationary phase and, depending 
on the duration of inflation, by the stimulatated emission from the protoinflationary relics. The conditions for the suppression or for the enhancement of protoinflationary curvature perturbations are related to the evolution of the pump fields controlling the transfer between the energy density of the background and the quasi-normal modes of the system.

The present paper is organized as follows. After introducing the governing equations,
in section \ref{sec2}, the quasi-normal mode of the system are derived 
in section \ref{sec3}. The fate of the large-scale 
curvature perturbations across the protoinflationary transition is investigated in 
section \ref{sec4} where the normalization 
of the fluctuations during the protoinflationary phase is also discussed. In section \ref{sec5} the nature of the 
transition is clarified in terms of the monotonicity properties of the pump fields 
accounting for the contribution of the inflaton and of the protoinflationary fluid to the total curvature perturbations. Explicit analytical and numerical examples are used to draw some general lessons on the dynamical features of the protoinflationary transition. Section \ref{sec6} contains some concluding remarks. The derivation of the coupled evolution equations of the quasi-normal modes related, asymptotically, to fluid phonons and inflaton quanta is reported in the appendix.

\renewcommand{\theequation}{2.\arabic{equation}}
\setcounter{equation}{0}
\section{Governing equations}
\label{sec2}
\subsection{General consideration}
The minimal set of assumptions characterizing the framework of the present investigation stipulates 
that the four-dimensional geometry is determined by the Einstein equations, supplemented by  the conservation equations accounting for the dynamics of  the inflaton and of the protoinflationary sources\footnote{Greek indices run from $0$ to $3$; the signature of the metric $g_{\alpha\beta}$ is mostly minus and $\nabla_{\alpha}$ 
denote the covariant derivative with respect to $g_{\alpha\beta}$.}:
\begin{eqnarray}
&& R_{\alpha}^{\beta} - \frac{1}{2} \delta_{\alpha}^{\beta} R = 
8\pi G \biggl[ T_{\alpha}^{\beta}(\varphi)
+ {\mathcal T}_{\alpha}^{\beta}(\rho_{pr}, p_{pr}) \biggr],
\label{g1}\\
&& g^{\alpha\beta} \nabla_{\alpha} \nabla_{\beta} \varphi + 
\frac{\partial V}{\partial\varphi} =0,   
\label{g2}\\
&& \nabla_{\alpha} {\mathcal T}_{\beta}^{\alpha} =0, \qquad g^{\alpha\beta} \, u_{\alpha}\, u_{\beta} = 1,
\label{g2a}
\end{eqnarray}
where $T_{\alpha}^{\beta}(\varphi)$ and ${\cal T}_{\alpha}^{\beta}(\rho_{pr},\, p_{pr})$ are, 
respectively, the energy-momentum tensors of the inflaton field $\varphi$  
and of the protoinflationary fluid:
\begin{eqnarray}
T_{\alpha}^{\beta}(\varphi) &=& \partial_{\alpha} \varphi \partial^{\beta} \varphi - \biggl[\frac{1}{2} g^{\mu\nu} \partial_{\mu} \varphi 
\partial_{\nu} \varphi - V(\varphi) \biggr] \,\delta_{\alpha}^{\beta},
\label{g3}\\
{\mathcal T}_{\alpha}^{\beta}(\rho_{pr},\, p_{pr})&=& (p_{pr} + \rho_{pr}) \,u_{\alpha} u^{\beta} - p_{pr} \,\delta_{\alpha}^{\beta}.
\label{g4}
\end{eqnarray}
The subscripts in the energy density and pressure remind of the protoinflationary origin of the fluid variables. In a conformally flat  background metric of 
the type $\overline{g}_{\alpha\beta} = a^2(\tau) \eta_{\alpha\beta}$ (where $a(\tau)$ is the scale factor in conformal time and $\eta_{\alpha\beta}$ is the Minkowski metric),  Eqs. (\ref{g1}), (\ref{g2}) and (\ref{g2a}) lead to a set of four equations 
\begin{eqnarray}
&& {\mathcal H}^2 = \frac{8\pi G}{3} \bigg[ a^2 \rho_{pr} + \frac{{\varphi'}^2}{2} + V\, a^2 \biggr], 
\label{FL1}\\
&& {\mathcal H}^2 - {\mathcal H}' = 4 \pi G [ a^2 (\rho_{pr} + p_{pr}) + {\varphi'}^2],
\label{FL2}\\
&& \varphi'' + 2 {\mathcal H} \varphi' + \frac{\partial V}{\partial \varphi} a^2 =0,
\label{FL3}\\
&& \rho_{pr}' + 3 {\mathcal H} ( \rho_{pr} + p_{pr}) =0,
\label{FL4}
\end{eqnarray}
which are not all independent and whose specific form is dictated by the fluid content of the primordial plasma. In Eqs. (\ref{FL1})--(\ref{FL4}) the prime 
denotes a derivation with respect to the conformal time coordinate $\tau$; 
furthermore  ${\mathcal H}= (\ln{a})^{\prime}$. The connection between ${\mathcal H}$ and the Hubble parameter is $ H= {\mathcal H}/a$. 
The effective energy and pressure densities of $\varphi$ are given by
\begin{equation}
\rho_{\varphi} = \frac{{\varphi'}^2}{2 a^2} + V(\varphi), \qquad
p_{\varphi} =    \frac{{\varphi'}^2}{2 a^2} - V(\varphi).
\label{FL5}
\end{equation}
Note that Eq. (\ref{FL3}) is equivalent to 
\begin{equation}
\rho_{\varphi}' + 3 {\mathcal H} (\rho_{\varphi} + p_{\varphi}) =0,\qquad c_{\varphi}^2 = \frac{p_{\varphi}'}{\rho_{\varphi}'}
= 1 + \frac{2 \,a^2}{3 {\mathcal H} \varphi'} \biggl(\frac{\partial V}{\partial \varphi}\biggr) \frac{{\mathcal H}}{\varphi'}.
\label{FL6}
\end{equation}

\subsection{Uniform curvature gauge} 
The most general scalar fluctuation of the four-dimensional metric is parametrized by four 
different functions whose number can be eventually reduced by specifying (either completely or partially) the coordinate system: 
\begin{equation}
 \delta_{\mathrm{s}} g_{00} = 2 a^2 \phi, \qquad \delta_{\mathrm{s}} g_{ij} = 2 a^2(\psi \delta_{ij} - \partial_{i} \partial_{j} E), \qquad 
 \delta_{\mathrm{s}} g_{0i} = - a^2  \partial_{i} B,
\label{UC0} 
\end{equation}
where $\delta_{\mathrm{s}}$ denotes the scalar mode of the corresponding tensor component; the full metric 
(i.e. background plus inhomogeneities) is given, in these notations, by $g_{\alpha\beta}(\vec{x}, \tau) = \overline{g}_{\alpha\beta}(\tau) + 
\delta_{\mathrm{s}} g_{\alpha\beta}(\vec{x},\tau)$ where, as already mentioned prior to Eqs. (\ref{FL1})--(\ref{FL4}) 
$\overline{g}_{\alpha\beta}(\tau) = a^2(\tau) \eta_{\alpha\beta}$.  For infinitesimal coordinate shifts  $\tau \to \overline{\tau} = \tau + \epsilon_{0}$ and 
$ {x}^{i} \to \overline{x}^{i} = x^{i} + \partial^{i}\epsilon$ the functions $\phi(\vec{x},\tau)$, $B(\vec{x},\tau)$, 
$\psi(\vec{x},\tau)$ and $E(\vec{x},\tau)$ introduced in Eq. (\ref{UC0}) transform as:
\begin{eqnarray}
&& \phi \to \overline{\phi} = \phi - {\cal H} \epsilon_0 - \epsilon_{0}' ,\qquad \psi \to \overline{\psi} = \psi + {\cal H} \epsilon_{0},
\label{phipsi}\\
&& B \to \overline{B} = B +\epsilon_{0} - \epsilon',\qquad E \to \overline{E} = E - \epsilon.
\label{EB}
\end{eqnarray}
In the uniform curvature gauge two out of the four functions of Eq. (\ref{UC0}) are set to zero \cite{hoff1}:
\begin{equation}
E=0, \qquad \psi =0.
\label{UC1}
\end{equation}
Starting from a gauge where $E$ and $\psi$ do not vanish, the perturbed line element can always be
brought in the form (\ref{UC1}) by demanding $\overline{E}=0$ and $\overline{\psi}=0$ in Eqs. (\ref{phipsi}) 
and (\ref{EB}). More specifically, if $E\neq 0$ and $\psi\neq 0$,  the uniform curvature gauge condition
can be recovered by fixing the gauge parameters as $\epsilon = E$ and $\epsilon_{0} = - \psi/{\mathcal H}$. This choice guarantees that, in the transformed coordinate system, $\overline{\psi}= \overline{E} =0$. 

The gauge condition of Eq. (\ref{UC1}) implies that the fluctuations of the spatial curvature vanish but, in this case, the perturbed metric also contains off-diagonal elements. With the condition (\ref{UC1}) the gauge freedom is totally fixed without the need of further conditions: because 
of this property the functions $\phi(\vec{x},\tau)$ and $B(\vec{x},\tau)$ bear an extremely simple 
relation to one of the conventional sets of gauge-invariant variables, as it will be shown later in this section. Finally, the off-diagonal coordinate system will prove very useful in section \ref{sec3} and in appendix \ref{APPA} for the analysis of the coupled system of quasi-normal modes. In the gauge (\ref{UC1}) the inhomogeneities of the energy-momentum tensors $T_{\alpha}^{\beta}(\varphi)$ and ${\cal T}_{\alpha}^{\beta}(\rho_{pr},\, p_{pr})$ are, respectively, 
\begin{eqnarray}
&& \delta_{\mathrm{s}}\,T_{0}^{0}= \frac{1}{a^2} \biggl( - \phi {\varphi'}^2 
+ \frac{\partial V}{\partial \varphi} a^2 \chi+ \chi' \varphi'\biggr),\qquad
\delta_{\mathrm{s}}\,T_{0}^{i} = - \frac{1}{a^2} \varphi' \partial^{i} \chi 
- \frac{{\varphi'}^2}{a^2} \partial^{i} B,
\label{UC3}\\
&&  \delta_{\mathrm{s}}\,T_{i}^{j}= \frac{1}{a^2}\biggl( \phi {\varphi'}^2 
+ \frac{\partial V}{\partial \varphi} a^2\chi - \chi' \varphi'\biggr) 
\delta_{i}^{j},
\label{UC4}\\
&& \delta_{\mathrm{s}} {\mathcal T}_{0}^{0} = \delta\rho_{pr}, \qquad \delta_{\mathrm{s}} {\mathcal T}_{i}^{j} = - \delta p_{pr}\, \delta_{i}^{j},\qquad \delta_{\mathrm{s}} {\mathcal T}_{0}^{i} = ( p_{pr} + \rho_{pr}) v^{i}.
\label{UC6}
\end{eqnarray}
where $\chi(\vec{x},\tau)$ and $v^{i}$ denote, respectively, the fluctuations of the inflaton $\varphi$ 
and the three-velocity field in the gauge (\ref{UC1}).
The fluctuations of the Einstein tensor ${\mathcal G}_{\mu}^{\nu} = R_{\mu}^{\nu} - R \delta_{\mu}^{\nu}/2$ in the gauge (\ref{UC1}) are instead:
\begin{eqnarray}
\delta_{\mathrm{s}} {\mathcal G}_{0}^{0} &=& \frac{2}{a^2} \biggl[
- {\mathcal H} \nabla^2 B  - 3 {\mathcal H}^2 \,\phi \biggr],
\label{UC7}\\
\delta_{\mathrm{s}} {\mathcal G}_{i}^{j} &=& \frac{1}{a^2} \biggl\{ \biggl[- 
2 ({\cal H}^2 + 2 {\cal H}') \phi - 2 {\cal H} \phi'\biggr]
- \nabla^2 \biggl( \phi + B' + 2 {\cal H} B \biggr)\biggr\} \delta_{i}^{j}
\nonumber\\
&+& \frac{1}{a^2}
\partial_{i}\partial^{j} \biggl(B' + 2 {\cal H}  B  +\phi\biggr),
\label{UC8}\\
\delta_{\mathrm{s}} {\mathcal G}_{0}^{i} &=&\frac{2}{a^2} \partial^{i}\biggl[ - {\cal H} \phi + ({\cal H}' - {\cal H}^2)B\biggr].
\label{UC9}
\end{eqnarray} 
The combination of Eqs. (\ref{UC3})--(\ref{UC4}) and (\ref{UC6}) with Eqs. (\ref{UC7}), (\ref{UC8}) and (\ref{UC9}) implies that the $(00)$ and $(0i)$ components of the perturbed Einstein equations with mixed indices become:
\begin{eqnarray}
&&{\mathcal H} \nabla^2 B + 3 {\mathcal H}^2 \phi = - 4\pi G a^2 \biggl(\delta\rho_{pr} + \delta\rho_{\varphi}\biggr),
\label{HG1}\\
&& ({\mathcal H}' - {\mathcal H}^2) \nabla^2 B - {\mathcal H} \nabla^2 \phi = 
4\pi G a^2 \biggl\{(p+ \rho) \theta_{pr} - \frac{1}{a^2}\biggl[ \varphi' \nabla^2 \chi + 
{\varphi'}^2 \nabla^2 B \biggr]\biggr\},
\label{HM2}
\end{eqnarray}
where $\theta_{pr}(\vec{x},\tau) =\partial_{i} \, v^{i}$. The variables $\delta\rho_{\varphi}$ and 
$\delta p_{\varphi}$ correspond to the fluctuations of the energy density and of the pressure of the inflaton field:
\begin{eqnarray}
&& \delta\rho_{\varphi} =  \frac{1}{a^2} \biggl(- \phi {\varphi'}^2 + \chi' \varphi' + a^2 
\frac{\partial V}{\partial \varphi} \chi\biggr),
\label{delta1}\\
&& \delta p_{\varphi} =  \frac{1}{a^2} \biggl(- \phi {\varphi'}^2 + \chi' \varphi' - a^2 
\frac{\partial V}{\partial \varphi} \chi\biggr).
\label{delta2}
\end{eqnarray}
To avoid lengthy notations we wrote $\delta \rho_{\varphi}$ (instead of $\delta_{\mathrm{s}} \rho_{\varphi}$),
 $\delta \rho_{pr}$ (instead of $\delta_{\mathrm{s}} \rho_{pr}$) and similarly for the corresponding pressures; this 
notation is fully justified and unambiguous once the scalar nature of the fluctuations has been established, as 
specified by the general formulae written above.  Bearing in mind this specification, the $(ij)$ component of the perturbed Einstein equations reads:
\begin{eqnarray}
&& \biggl\{ \biggr[ - ({\mathcal H^2} + 2 {\mathcal H}') \phi - {\mathcal H} \phi' - 
\frac{1}{2} \nabla^2 ( \phi + B' + 2 {\mathcal H} B) \biggr] \delta_{i}^{j} 
\nonumber\\
&&+ \frac{1}{2} \partial_{i}\partial^{j} \biggl[ \phi + B' + 2 {\mathcal H} B \biggr] \biggr\} = 4\pi G a^2 \biggl\{ - \biggl[ \delta p_{pr} \, + 
\delta p_{\varphi} \biggr] \delta_{i}^{j} + \Pi_{i}^{j} \biggr\}.
\label{Hij}
\end{eqnarray}
The separation of the traceless part from the trace in Eq. (\ref{Hij}) leads to two independent relations:
\begin{eqnarray}
&& ({\mathcal H}^2 + 2 {\mathcal H}') \phi + {\mathcal H} \phi' + \frac{1}{3} \nabla^2(\phi + B' + 2 {\mathcal H} B) =
4\pi G a^2 (\delta p_{pr} + \delta p_{\varphi}),
\label{sep1}\\
&& \partial_{i}\partial^{j} [ \phi + B' + 2 {\mathcal H} B] - \frac{1}{3} \nabla^2 [ \phi + B' + 2 {\mathcal H} B] \delta_{i}^{j}= 
8\pi G a^2 \Pi_{i}^{j}.
\label{sep2}
\end{eqnarray}
If the anisotropic stresses $\Pi_{i}^{j}$ is neglected, 
Eq. (\ref{Hij}) is the equivalent to the following pair of conditions:
\begin{eqnarray}
&&  ({\mathcal H}^2 + 2 {\mathcal H}') \phi + {\mathcal H} \phi'  = 4\pi G a^2 (\delta p_{pr} + \delta p_{\varphi}),
\label{sep3}\\
&& \phi + B' + 2 {\mathcal H} B =0.
\label{sep4}
\end{eqnarray}
The evolution equation for the perturbed inflaton is:
\begin{eqnarray}
\chi'' + 2 {\mathcal H} \chi' - \nabla^2 \chi + 
\frac{\partial^2 V}{\partial\varphi^2} a^2 \chi + 2 \phi \frac{\partial V}{\partial\varphi} a^2 - 
\varphi' \phi' - \varphi' \nabla^2 B=0.
\label{chi}
\end{eqnarray}
Finally, by perturbing the covariant conservation equation of the fluid energy-momentum tensor the evolution equation 
for the density fluctuation is:
\begin{equation}
\delta\rho_{pr}^{\,\prime} + (p + \rho) \theta_{pr} + 3 {\mathcal H}(\delta p_{pr} + \delta \rho_{pr}) =0,
\label{dr}
\end{equation}
while the equation for the three-divergence of the velocity becomes:
\begin{equation}
( \theta_{pr} + \nabla^2 B)' + \frac{[ p_{pr}' + {\mathcal H} ( p_{pr} + \rho_{pr})] }{( p_{pr} + \rho_{pr})}( \theta_{pr} + \nabla^2 B) 
+\frac{\nabla^2 \delta p_{pr}}{ (p_{pr} + \rho_{pr}) } +  \nabla^2 \phi =0.
\label{th}
\end{equation}
Equations (\ref{chi}) and (\ref{th}) represent the staring point for the derivation of the quasi-normal modes of 
the system. 
\subsection{Gauge-invariant observables}
 The coordinate system defined by Eq. (\ref{UC1}) completely fixes the gauge freedom without the need 
 of further subsidiary conditions. The absence of spurious gauge modes is then guaranteed as it happens 
 for other choices of coordinates removing completely the gauge freedom such as the conformally 
 Newtonian gauge. As a consequence,
  the perturbation variables defined in the gauge (\ref{UC1}) bear a simple relation 
to the various gauge-invariant observables. The specific connection between the degrees of freedom 
defined in the uniform curvature gauge and other common gauge-invariant combinations
will now be outlined. 

The curvature perturbation on comoving orthogonal hypersurfaces (conventionally denoted by ${\mathcal R}$) 
and the Bardeen potential (conventionally denoted by $\Psi$)
coincide, up to time-dependent functions, with $\phi(\vec{x},\tau)$ and $B(\vec{x},\tau)$:
\begin{equation}
\Psi = - {\mathcal H} B,\qquad {\mathcal R} = - \frac{{\mathcal H}^2}{{\mathcal H}^2 - {\mathcal H}'} \phi.
\label{G1}
\end{equation}
The result of Eq. (\ref{G1}) can be easily derived from the explicit gauge transformation 
relating the uniform curvature hypersurfaces with the comoving orthogonal 
hyepersurfaces. Conversely, from the customary expression of ${\mathcal R}$ in terms 
of the gauge-invariant Bardeen potentials $\Phi$ and $\Psi$ the result of Eq. (\ref{G1}) can be cross-checked.
The curvature perturbations on comoving orthogonal hypersurfaces are given by:
\begin{equation}
{\mathcal R} = - \Psi - \frac{{\mathcal H} ( {\mathcal H} \Phi + \Psi')}{{\mathcal H}^2 - {\mathcal H}'}.
\label{G2}
\end{equation}
But in terms of the variables introduced in Eq. (\ref{UC0}) the expression of $\Phi$ and $\Psi$ is:
\begin{equation}
 \Phi = \phi + ( B - E') ' + {\cal H}( B- E'),\qquad \Psi = \psi - {\cal H} ( B - E').
\label{G3}
\end{equation}
According to Eq. (\ref{G3}),  in the gauge (\ref{UC1}) $\Phi = \phi + B' + {\mathcal H} B$ and $\Psi = - {\mathcal H} B$. By inserting into Eq. (\ref{G2}) the expressions for $\Phi$ and $\Psi$ written in the gauge (\ref{UC1}) 
the results of Eq. (\ref{G1}) are immediately recovered.
The total density contrast on uniform curvature hypersurfaces can be expressed as 
\begin{equation}
\zeta = - \frac{\delta\rho_{\mathrm{t}}}{\rho_{\mathrm{t}}'} {\mathcal H}, \qquad \delta \rho_{\mathrm{t}} = (\delta \rho_{pr} + \delta\rho_{\varphi} ),\qquad \rho_{\mathrm{t}} = \rho_{\varphi} + \rho_{pr}.
\label{G4}
\end{equation}
From Eq. (\ref{HG1}) recalling Eqs. (\ref{G1}) and (\ref{G4}) we can also 
obtain the relation between ${\mathcal R}$, $\zeta$ and $\Psi$:
\begin{equation}
\zeta = {\mathcal R} + \frac{\nabla^2 \Psi}{12 \pi G a^2 (p_{\mathrm{t}} + \rho_{\mathrm{t}})}.
\label{G5}
\end{equation}
It is relevant to remind that $\zeta$ and ${\mathcal R}$ are often used interchangeably. This is justified 
provided the wavelengths under considerations are sufficiently larger than the Hubble radius at the corresponding time. Otherwise the two variables $\zeta$ and ${\mathcal R}$ are physically different. 
\renewcommand{\theequation}{3.\arabic{equation}}
\setcounter{equation}{0}
\section{Quasi-normal modes of the system}
\label{sec3}
The system of section \ref{sec2}
 describing the evolution across the protoinflationary boundary has two 
 asymptotic limits corresponding to the situation where one of the two components is either absent or dynamically negligible. If the protonflationary fluid and the inflaton are simultaneously present the evolution is characterized by a pair of (interacting) quasi-normal modes which are the generalization of the normal modes obtainable in the case of a single component. After swiftly summarizing what happens in the two asymptotic limits, 
the derivation of the quasi-normal modes of the whole system will be presented. The interested reader may also consult 
appendix \ref{APPA} where some of the technical results involved in the derivation are collected.

In the limit
$\varphi'(\tau) \to 0$ and $\chi(\vec{x},\tau) \to 0$ the fluctuations 
of the inflaton energy density and of the inflaton pressure are both vanishing, i.e. $\delta \rho_{\varphi} = \delta p_{\varphi}=0$. Since $\delta p_{pr} = c_{\mathrm{s}}^2 \delta \rho_{pr}$,  Eq. (\ref{HG1}) can be multiplied by $c_{\mathrm{s}}^2$ and summed  to Eq. (\ref{sep3}). After simple algebra the following result will be obtained:
\begin{equation}
{\mathcal H} \phi' + [ {\mathcal H}^2 ( 1 + 3 c_{\mathrm{s}}^2 ) + 2 {\mathcal H}'] \phi = -
 {\mathcal H} c_{\mathrm{s}}^2 \nabla^2 B.
\label{QN1}
\end{equation}
Introducing the variables $\alpha$ and $\gamma$ defined as\footnote{The background-dependent functions $\alpha^2$ and $\gamma$ mentioned in Eqs. (\ref{QN1})--(\ref{QN2}) illustrate, for convenience, the notations Refs. \cite{lukash,strokov}. In the rest of the paper it will be more practical to adopt a slightly different set of variables which are introduced in Eqs. (\ref{QN4}) and (\ref{QN5}).}: 
\begin{equation}
\gamma = \frac{3}{2} \biggl( 1 + \frac{p_{pr}}{\rho_{pr}}\biggr),\qquad \alpha^2 = \frac{\gamma}{4\pi G c_{\mathrm{s}}^2},
\label{QN2}
\end{equation}
Eq. (\ref{QN1}) becomes
\begin{equation}
\biggl(\frac{\phi}{\gamma}\biggr)' = -  \frac{\nabla^2 B}{4 \pi G \alpha^2}.
\label{QN3}
\end{equation}
Differentiating both sides of Eq. (\ref{QN3}) with respect to $\tau$, two kinds of terms (proportional to $\nabla^2 B$ and to $\nabla^2 B'$) will arise; using then Eq. (\ref{sep3}) to eliminate the terms proportional to $\nabla^2B'$ 
and Eq. (\ref{QN3}) to eliminate the terms containing $\nabla^2B$, a decoupled equation for $\phi$ is readily obtained. Recalling the simple relation between $\phi$ and ${\mathcal R}$ mentioned in (\ref{G1}) the resulting equation becomes:
\begin{equation}
{\mathcal R}_{pr}'' + 2 \frac{z_{pr}'}{z_{pr}} {\mathcal R}_{pr}' - c_{\mathrm{s}}^2 \nabla^2 {\mathcal R}_{pr} =0, 
\label{QN4}
\end{equation}
where ${\mathcal R}_{pr}$ is the curvature perturbation on comoving orthogonal hypersurfaces and 
\begin{equation}
{\mathcal R}_{pr} = - \frac{\phi}{\gamma}= - \frac{q_{v}}{z_{pr}},
\qquad z_{pr} = \alpha \, a= 
\frac{a^2 \sqrt{p_{pr} + \rho_{pr}}}{{\mathcal H} \, c_{\mathrm{s}}}.
\label{QN5}
\end{equation}
The function $q_{v}$ introduced in Eq. (\ref{QN5}) is actually the normal mode of the system obeying:
\begin{equation}
q_{v}'' - c_{\mathrm{s}}^2 \nabla^2 q_{v} - \frac{z_{pr}''}{z_{pr}} q_{v} =0.
\label{QN5a}
\end{equation}
By using the momentum constraint in the purely hydrodynamical case (i.e. $\varphi'(\tau)\to 0$ and $\chi(\vec{x},\tau) \to 0$ in Eq. (\ref{HM2})) the following chain of equalities holds 
\begin{equation}
\nabla^2{\mathcal R}_{pr} = - \nabla^2 \biggl(\frac{\phi}{\gamma}\biggr) =  \frac{4 \pi G a^2 (p_{pr} + \rho_{pr})}{\gamma\,\,{\mathcal H}} (\theta_{pr} + \nabla^2 B),
\label{QN7}
\end{equation}
which also implies, always neglecting the inflaton, that $ \nabla^2{\mathcal R}_{pr} = {\mathcal H}  (\theta_{pr} + \nabla^2 B)$.
The same steps leading to Eqs. (\ref{QN4}) and (\ref{QN5}) can be applied to the case of scalar field matter
in the absence of protoinflationary component (i.e. $\rho_{pr} = p_{pr} = 0$ and $\delta\rho_{pr} = \delta p_{pr}=
\theta_{pr} =0$). The evolution equation of the curvature 
perturbation will be given, in this case, by:
\begin{equation}
{\mathcal R}_{\varphi}'' + 2 \frac{z_{\varphi}'}{z_{\varphi}} {\mathcal R}_{\varphi}' - \nabla^2 {\mathcal R}_{\varphi} =0, \qquad z_{\varphi} =
\frac{a \varphi'}{{\mathcal H}}.
\label{QN6}
\end{equation}
The separate use of the momentum constraint in the scalar field case  (i.e. $\theta_{pr}=0$ in Eq. (\ref{HM2})) 
leads to the analog of Eq. (\ref{QN7}):
\begin{equation}
{\mathcal R}_{\varphi} = - \frac{4 \pi G \, \varphi'}{{\mathcal H}} \chi = - \frac{q_{\chi}}{z_{\varphi}},
\label{QN8}
\end{equation}
where $q_{\chi} = a \chi \equiv - z_{\varphi} \, {\mathcal R}_{\varphi}$ obeys, from Eq. (\ref{QN6}), the following equation:
\begin{equation}
q_{\chi}'' - \nabla^2 q_{\chi} - \frac{z_{\varphi}''}{z_{\varphi}} q_{\chi}=0,
\label{QN8a}
\end{equation}
which is the analog of Eq. (\ref{QN5a}) holding in the absence of inflaton contribution. 
When dealing with the process of parametric amplification, the variables $z_{pr}(t)$ and $z_{\varphi}(t)$ are dubbed 
pump fields since they control the rate of energy transfer from the background to the fluctuations. 
This terminology is often used in quantum optics \cite{QO} (see also \cite{six}) and shall also 
be employed in the forthcoming considerations. 

The results obtained in 
Eqs. (\ref{QN4})--(\ref{QN7}) assume the absence of the inflaton field. Conversely 
Eqs. (\ref{QN6}) and (\ref{QN8}) assume the absence of the protoinflationary fluid. 
If the contributions of the fluid and of the scalar field are simultaneously taken into account,  the evolution equations 
of the resulting system can be reduced to a pair of coupled equations whose solution gives directly 
the curvature fluctuations on comoving orthogonal hypersurfaces. Indeed, by keeping both contributions,  Eq. (\ref{HM2}) implies
\begin{equation}
\phi = \frac{4\pi G \, \varphi'}{{\mathcal H}} \chi + \frac{4 \pi G a^2 (p_{pr} + \rho_{pr})}{{\mathcal H}} u,
\label{QN9}
\end{equation}
where the notation $(\theta_{pr} + \nabla^2\,B) = - \nabla^2 u$ has been introduced. To simplify the discussion, we shall assume that the protoinflationary fluid is characterized by a constant barotropic index $w$ 
so that $c_{\mathrm{s}} = \sqrt{w}$. The derivation of the coupled system of the quasi-normal modes is reported in the appendix \ref{APPA}.   The final equations obeyed by $q_{v}$ and $q_{\chi}$ are:
\begin{eqnarray}
&& q_{v}^{\prime\prime} - c_{\mathrm{s}}^2 \nabla^2 q_{v} + {\mathcal A}_{v\,v}(\tau) q_{v} + {\mathcal B}_{v\,\chi}(\tau) q_{\chi} + {\mathcal C}_{v\,\chi}(\tau) q_{\chi}^{\prime} =0,
\label{NMM1}\\
&& q_{\chi}^{\prime\prime} - \nabla^2 q_{\chi} + \overline{{\mathcal A}}_{\chi\,\chi}(\tau) q_{\chi} + \overline{{\mathcal B}}_{\chi\,v}(\tau) q_{v} + \overline{{\mathcal C}}_{\chi\,v}(\tau) q_{v}^{\prime} =0.
\label{NMM2}
\end{eqnarray}
The coefficients ${\mathcal A}_{v\,v}(\tau)$, ${\mathcal B}_{v\,\chi}(\tau)$ and ${\mathcal C}_{v\,\chi}(\tau)$ depend 
on the conformal time coordinate and are given by the following expressions\footnote{In Eqs. (\ref{A1})--(\ref{C1}) and (\ref{A2})--(\ref{C2}) natural Planckian units $\overline{M}_{\mathrm{P}} =1$ are used. The same units are also used in the second part of appendix \ref{APPA} where the explicit derivation of Eqs. (\ref{NMM1}) and (\ref{NMM2}) is reported.}:
\begin{eqnarray}
{\mathcal A}_{v\,v}(\tau) &=& - \frac{(3 w -1)}{2} {\mathcal H}'  - \frac{(3w -1)^2}{4} {\mathcal H}^2 
- \frac{a^2 (p_{pr} + \rho_{pr})}{4 {\mathcal H}^2 }\biggl\{ {\varphi'}^2 ( w -1) 
\nonumber\\
&-& 2 \biggl[ {\mathcal H}^2 (3 w +1) + 2 {\mathcal H}'\biggr]\biggr\},
\label{A1}\\
{\mathcal B}_{v\,\chi}(\tau) &=& - \frac{a \sqrt{p_{pr} + \rho_{pr}}}{4 {\mathcal H} \,\sqrt{w}}\biggl\{ \biggl(\frac{\varphi'}{{\mathcal H}}\biggr) \biggl[{\varphi'}^2 ( w -1) - 2 ( {\mathcal H}^2 (3 w +1) + 2 {\mathcal H}')\biggr]
\nonumber\\
 &-& 2 a^2 (w +1) \frac{\partial V}{\partial \varphi}  + 2 {\mathcal H} \varphi' (1 - w)\biggr\},
 \label{B1}\\
 {\mathcal C}_{v\,\chi}(\tau) &=& \frac{a \sqrt{p + \rho}}{2 {\mathcal H}\,\sqrt{w}} \varphi' ( 1 - w). 
 \label{C1}
\end{eqnarray} 
The coefficients $\overline{{\mathcal A}}_{\chi\,\chi}(\tau)$, $\overline{{\mathcal B}}_{\chi\,v}(\tau)$ and 
$\overline{{\mathcal C}}_{\chi\,v}(\tau)$ are instead:
\begin{eqnarray}
\overline{{\mathcal A}}_{\chi\,\chi}(\tau) &=& a^2 \frac{\partial^2 V}{\partial \varphi^2}  - {\mathcal H}^2 - {\mathcal H}' + 
 \frac{\varphi'}{4 {\mathcal H}}\biggl\{ 8 a^2 \frac{\partial V}{\partial \varphi} 
+ \frac{\varphi'}{{\mathcal H}}\biggl[4 ({\mathcal H}' + 2 {\mathcal H}^2) 
\nonumber\\
&+& a^2 ( p_{pr} + \rho_{pr}) \biggl(\frac{w -1}{w} \biggr) 
 \biggr]\biggr\},
 \label{A2}\\
 \overline{{\mathcal B}}_{\chi\,v}(\tau) &=& \frac{a \sqrt{p_{pr} + \rho_{pr}}}{4 {\mathcal H}} \biggl\{ \frac{(3 w -1)\,(w -1)}{w}\varphi' {\mathcal H} + 4 a^2 \frac{\partial V}{\partial \varphi} 
 \nonumber\\
&+& \frac{\varphi'}{{\mathcal H}} \biggl[ 4 ({\mathcal H}' + 2 {\mathcal H}^2)  + a^2 (p_{pr} + \rho_{pr}) \biggl(\frac{w- 1}{\sqrt{w}} \biggr)\biggr]\biggr\},
 \label{B2}\\
 \overline{{\mathcal C}}_{\chi\,v}(\tau) &=&- \frac{a \sqrt{p_{pr} + \rho_{pr}}}{2 {\mathcal H}} \varphi' \biggl(\frac{w -1}{w} \biggr)\,\sqrt{w}.
\label{C2}
\end{eqnarray}
In the limit $\rho_{pr}\to 0$ and $p_{pr} \to 0$, the equation for $q_{\chi}$ coincides with 
the equation obeyed by $z_{\varphi} \,{\mathcal R}_{\varphi}$ and reported in Eq. (\ref{QN8a}); this is not a surprise 
since, in the gauge (\ref{UC1}) and when $\rho_{pr} =0$, $q_{\chi} = - z_{\varphi} \, {\mathcal R}_{\varphi}$. 
In the opposite limit, i.e.  $\varphi^{\prime} \to 0$,  the equation obeyed by $q_{v}$ 
coincides with the equation obeyed by $z_{pr} \,{\mathcal R}_{pr}$ (in the case $c_{\mathrm{s}}^2 = w$) and reported in Eq. (\ref{QN5a}). In the general case, using the notations developed so far, the total curvature perturbations can be written as
\begin{equation}
\phi = \frac{z_{pr}}{2 a^2}\, c_{\mathrm{s}}^2\,q_{v} + \frac{z_{\varphi}}{2 a^2} \, q_{\chi}.
\label{C3}
\end{equation}
From equation (\ref{C3}) the curvature perturbations can also be computed and they are: 
\begin{equation}
{\mathcal R} = - \frac{z_{pr} \, c_{\mathrm{s}}^2}{z_{pr}^2 c_{\mathrm{s}}^2 + z_{\varphi}^2} q_{v} - \frac{z_{\varphi}}{z_{pr}^2 c_{\mathrm{s}}^2 + z_{\varphi}^2} q_{\chi}.
\label{rgen}
\end{equation}
It is immediate to show from Eqs. (\ref{QN5}) and (\ref{rgen}) that ${\mathcal R} \to {\mathcal R}_{pr}$ when $z_{\varphi} \to 0$. In the same way when $z_{pr} \to 0$ Eqs. (\ref{QN8}) and (\ref{rgen}) imply that ${\mathcal R} \to {\mathcal R}_{\varphi}$.
The quasi-normal modes $q_{v}$ and $q_{\chi}$ describe, asymptotically, the excitations 
corresponding to fluid phonons and inflaton quanta.

\renewcommand{\theequation}{4.\arabic{equation}}
\setcounter{equation}{0}
\section{Across the protoinflationary transition}
\label{sec4}
The crudest model for the transition stipulates that prior to the onset 
of inflation the perfect and irrotational fluid dominates the total energy density and pressure. In a slightly inhomogeneous space-time the energy-momentum tensor experiences a finite discontinuity on the hypersurface of constant energy density when the slow-roll dynamics starts off. 
This choice for the matching hypersurface is adopted in the case 
of post-inflationary transitions \cite{trans} and it is interesting to scrutinize its implications in describing the protoinflationary boundary. 
The logic of the sudden approximation is to assume that $z_{\varphi}(t) \to 0$
during the protoinflationary phase. Conversely during the slow-roll epoch $z_{pr}(t) \to 0$. The sudden approximation (together with the 
required continuity of the extrinsic curvature) implies the suppression of the density contrast and of the metric fluctuation across the protoinflationary boundary.
The potential limitations of the sudden approximation are scrutinized in section \ref{sec5} where the protoinflationary dynamics 
is described in terms of a class of exact solutions of the background equations which will be presented later.

\subsection{The continuity of the extrinsic curvature}
If the stress tensor undergoes a finite discontinuity on a space-like hypersurface the inhomogeneities are matched by requiring the continuity of the induced three metric and of the extrinsic curvature on that hypersurface.  The extrinsic curvature is defined as
\begin{equation}
\overline{K}_{ij} = \frac{1}{2N} \biggl[ \nabla_{i} N_{j} + \nabla_{j} N_{i} - \gamma_{ij}^{\,\,\prime} \biggr].
\label{EX1}
\end{equation}
Recalling Eq. (\ref{UC0}), in a generic coordinate system the lapse function, the shift vectors and three-metric $\gamma_{ij}$ are, respectively:
\begin{equation}
N_{i} = a^2 \partial_{i} \overline{B}, \qquad N^2 = (1 + 2 \overline{\phi}) a^2, \qquad \gamma_{ij} = a^2( 1 - 2 \overline{\psi}) \delta_{ij} + 
2 a^2 \partial_{i} \partial_{j} \overline{E}.
\label{EX2}
\end{equation}
Using Eq. (\ref{EX2}) into Eq. (\ref{EX1}) the covariant and mixed 
components of the extrinsic curvature read, to first order in the scalar metric perturbations, 
\begin{eqnarray}
\overline{K}_{ij}(\vec{x},\tau) &=& - a {\mathcal H} \delta_{i j}  + a \biggl[ \partial_{i} \partial_{j}(\overline{B} - \overline{E}' - 2 {\mathcal H} \overline{E}) + (\overline{\psi}' + 2 {\mathcal H} \overline{\psi} + {\mathcal H}\overline{\phi})\delta_{ij} \biggr], 
\nonumber\\
\overline{K}_{i}^{j}(\vec{x},\tau) &=& -\frac{1}{a} \biggl[ \delta_{i}^{j} \biggl( {\mathcal H} \overline{\phi} + \overline{\psi} ' \biggr) + \partial_{i}\partial^{j}( \overline{B} - \overline{E}') \biggr].
\label{EX3}
\end{eqnarray}
The continuity of the background extrinsic 
curvature implies that across the transition the scale factor and the Hubble 
rate must be continuous. In cosmic time,  a continuous form of the scale factor can be written when, for instance, the inflationary phase 
is characterized by a set of constant slow-roll parameters (see, e.g. Eq. (\ref{srpar}) of section \ref{sec4} for a general definition of the slow roll parameters). In this case we shall have that:
\begin{eqnarray}
a_{pr}(t) &=& a_{*}\biggl(\frac{t}{t_{*}}\biggr)^{\alpha}, \qquad t \leq t_{*},
\nonumber\\
a_{inf}(t) &=& a_{*} \biggl[\frac{\alpha}{\beta} \biggl(\frac{t}{t_{*}}\biggr) + \frac{\beta - \alpha}{\beta} \biggr]^{\beta}, 
\qquad t> t_{*}.
\label{EX6}
\end{eqnarray}
In Eq. (\ref{EX6}) the inflationary evolution is realized for $\beta \gg 1$. For some applications 
it is useful to recall the conformal time parametrization where the continuity of the scale factors across the protoinflationary boundary can be expressed as:
\begin{eqnarray}
a_{pr}(\tau) &=& a_{*}\biggl[\biggl(\frac{\overline{\alpha}}{\overline{\beta}} + 1 \biggr) \frac{\tau_{*}}{\tau_{1}} + \frac{\tau}{\tau_{1}}\biggr]^{\overline{\alpha}}, \qquad - \biggl(\frac{\overline{\alpha}}{\overline{\beta}} + 1 \biggr) \tau_{*}< \tau \leq - \tau_{*},
\nonumber\\
a_{inf}(\tau) &=&\bigg(\frac{\overline{\alpha}}{\overline{\beta}} \, \frac{\tau_{*}}{\tau_{1}} \biggr)^{\overline{\alpha}} \biggl( - \frac{\tau_{*}}{\tau} \biggr)^{\overline{\beta}},\qquad - \tau_{*}\leq \tau \leq - \tau_{1},
\label{EX7}
\end{eqnarray}
where $\overline{\alpha} = \alpha/(1 - \alpha)$ and $\overline{\beta} = \beta/(\beta -1)$. From Eq. 
(\ref{EX6}) the scale factor and its first derivative are continuous in $t_{*}$. By going in the conformal 
parametrization $ a(\tau)\,d\tau = d t$ the time coordinate $\tau$ becomes negative and therefore 
the conformal time scale factor and its first derivative with respect to $\tau$  are continuous in $-\tau_{*}$.
In the parametrization of Eq. (\ref{EX7}) the inflationary regime occurs for $\beta \gg 1$ and $\overline{\beta} \to 1$.

From Eq. (\ref{EX3}) the continuity of the inhomogeneous part of $\overline{K}_{i}^{j}$ and $\overline{K}_{ij}$ 
implies the separate continuity of the combinations 
\begin{equation}
[\,\overline{\psi} \,]_{\pm} =0,\qquad [\,\overline{E} \,]_{\pm} =0,\qquad 
[\,{\mathcal H} \overline{\phi} + \overline{\psi}'\,]_{\pm} =0, \qquad [\,\overline{B} - \overline{E}'\,]_{\pm} =0,
\label{EX7a}
\end{equation}
where, following the general treatment of sudden transitions \cite{trans}, the subscript $\pm$ 
denotes the jump of the corresponding quantity across the transition (i.e. $[\,f\,]_{\pm} = f_{+} - f_{-}$).
In the coordinate system where $\overline{\tau}$ is constant, the equation for the hypersurface of constant 
energy density  becomes $\overline{\delta\rho}_{\mathrm{t}} =0$. But since 
\begin{equation}
\delta \rho_{\mathrm{t}} \to \overline{\delta\rho}_{\mathrm{t}} = \delta \rho_{\mathrm{t}} - \rho_{\mathrm{t}}' \epsilon_{0}, \qquad \overline{\delta\rho}_{\mathrm{t}} = \overline{\delta\rho_{pr}} + \overline{\delta\rho_{\varphi}},
\label{EX4}
\end{equation}
the condition $\overline{\delta\rho}_{\mathrm{t}} =0$  implies $\epsilon_{0} =  \delta \rho_{\mathrm{t}}/\rho_{\mathrm{t}}'$.
Recalling now the expressions for $\overline{\phi}$, $\overline{\psi}$, $\overline{E}$ and $\overline{B}$ 
stemming from Eqs. (\ref{phipsi}) and (\ref{EB}), the conditions of Eq. (\ref{EX7a}) become 
\begin{eqnarray}
&& \biggl[\,\psi + {\mathcal H} \frac{\delta \rho_{\mathrm{t}}}{\rho_{\mathrm{t}}'}\,\biggr]_{\pm} =0,\qquad 
\biggl[\,B- E' +  \frac{\delta \rho_{\mathrm{t}}}{\rho_{\mathrm{t}}'}\,\biggr]_{\pm} =0,
\nonumber\\
&& \biggl[\,{\mathcal H}\phi + \psi' + ({\mathcal H}' - {\mathcal H}^2) \frac{\delta \rho_{\mathrm{t}}}{\rho_{\mathrm{t}}'} \,\biggr]_{\pm} =0.
\label{EX7b}
\end{eqnarray}
Equations (\ref{EX7b}) are general and can be studied in any gauge. According to Eq. (\ref{EX7b}), 
the continuity of the extrinsic curvature 
in the gauge (\ref{UC1}) implies that the following combinations must separately be continuous 
\begin{eqnarray}
&& \biggl[ B + \frac{\delta\rho_{\mathrm{t}}}{\rho_{\mathrm{t}}'} \biggr]_{\pm} =0,\qquad  \biggl[\frac{\delta\rho_{\mathrm{t}}}{\rho_{\mathrm{t}}'} \biggr]_{\pm}=0,
\label{EX8}\\
&& \biggl[{\mathcal H} \phi + ({\mathcal H}' - {\mathcal H}^2) \frac{\delta\rho_{\mathrm{t}}}{\rho_{\mathrm{t}}'} \biggr]_{\pm} =0.
\label{EX9}
\end{eqnarray}
Since ${\mathcal H}$ is continuous, Eq. (\ref{EX9}) reduces to 
\begin{equation}
\biggl[{\mathcal H} \phi + {\mathcal H}'  \frac{\delta\rho_{\mathrm{t}}}{\rho_{\mathrm{t}}'} \biggr]_{\pm} =0.
\label{EX10}
\end{equation}
The Hamiltonian constraint of Eq. (\ref{HG1}) can be written in the form:
\begin{equation}
\delta \rho_{\mathrm{t}} = - \frac{{\mathcal H}}{4\pi G a^2} [ \nabla^2 B + 3 {\mathcal H} \phi],
\label{EX11}
\end{equation}
Equations (\ref{EX8}) and (\ref{EX10}) are then equivalent to the two conditions:
\begin{equation}
[\, B\,]_{\pm} =0, \qquad \biggl[- \frac{{\mathcal H}^2}{{\mathcal H}^2 - {\mathcal H}'} \phi \biggr]_{\pm} =0.
\label{EX12}
\end{equation}
But according to Eq. (\ref{G1}) we have that $\Psi = - {\mathcal H} B$ and ${\mathcal R} = - {\mathcal H}^2/({\mathcal H}^2 - {\mathcal H}') \phi$; thus the continuity of the scale factor and of ${\mathcal H}$ implies the continuity of ${\mathcal R}$ and $\Psi$ across the protoinflationary transition. 
The evolution will now be separately solved during the protoinflationary phase and during the inflationary phase. 
The mathching conditions expressed by Eq. (\ref{EX12}) agree with former 
treatments \cite{trans} but in the coordinate system defined by Eq. (\ref{UC1}). 

\subsection{Protoinflationary evolution}
During the protoinflationary phase and for $z_{\varphi} \to 0$ the system reduces to the triplet of equations
\begin{eqnarray}
&& \dot{B} + 2 H\,B + \biggl(\frac{\dot{H}}{H^2}\biggr) \frac{{\mathcal R}}{a} =0,
\label{SU1}\\
&& \dot{{\mathcal R}} = - \biggl(\frac{H^2}{\dot{H}}\biggr) \frac{c_{\mathrm{s}}^2}{a} \nabla^2 B,
\label{SU2}\\
&& \delta_{\mathrm{t}} = -\frac{2}{3} \frac{\nabla^2 B}{H a} - 2 \frac{\dot{H}}{H^2} {\mathcal R}, \qquad \delta_{\mathrm{t}} = \frac{\delta\rho_{pr}}{\rho_{pr}},
\label{SU3}
\end{eqnarray}
where the overdot denotes a derivation with respect to the 
cosmic time coordinate and  $\delta_{\mathrm{t}}$ is the total density contrast which is dominated, in this case, by the protoinflationary fluid.  Equation (\ref{SU1}) comes from Eq. (\ref{sep4}); Eq. (\ref{SU2}) is Eq. (\ref{QN1}) but written in the cosmic time coordinate; Eq. (\ref{SU3}) derives from Eq. (\ref{HG1}).

To enforce a correct normalization on the solutions we proceed as follows. 
After promoting the normal mode  $q_{v}(\vec{x},\tau)$ and its conjugate momentum to the status of field operators obeying 
canonical commutation relations at equal times, the mode expansion for $\hat{q}_{v}(\vec{x},\tau)$ becomes
\begin{equation}
\hat{q}_{v}(\vec{x},\tau) = \frac{1}{(2 \pi)^{3/2}} \int d^{3}k \biggl[ f_{k}(\tau) \, \hat{a}_{\vec{k}} \, e^{- i \, \vec{k}\cdot \vec{x}} 
+ f_{k}^{*}(\tau) \, \hat{a}_{\vec{k}}^{\dagger} \, e^{ i \, \vec{k}\cdot \vec{x}} \biggr],
\label{Q1}
\end{equation}
where $[\hat{a}_{\vec{k}}, \, \hat{a}_{\vec{p}}^{\dagger}] = \delta^{(3)}(\vec{k} - \vec{p})$ and the mode function $f_{k}(\tau)$ 
obeys 
\begin{eqnarray}
&& f_{k}'' + \biggl[ k^2 \, c_{\mathrm{s}}^2 - \frac{z_{pr}''}{z_{pr}}\biggr] f_{k} =0,\qquad z_{pr} = \frac{a^2 \sqrt{p_{pr} + \rho_{pr}}}{{\mathcal H} c_{\mathrm{s}}},
\label{Q2}\\
&& \frac{z_{pr}''}{z_{pr}} = \frac{\nu^2 -1/4}{y^2}, \qquad y= \biggl(\frac{\overline{\alpha}}{\overline{\beta}} + 1\biggr) \tau_{*} + \tau, \qquad \nu = \frac{3 ( 1 - w)}{2 ( 3 w + 1)}.
\label{Q3}
\end{eqnarray}
In the case $w = 1/3$ (corresponding to a radiation fluid) $\nu = 1/2$; it is interesting to remark that, in the case 
$w =0$ (but with $c_{\mathrm{s}} \neq 0$), $\nu = 3/2$ allowing for a flat spectrum of phonons, as discussed in \cite{lukash}. Equation 
(\ref{Q2}) can be solved exactly in terms of Hankel functions and the solution is:
\begin{equation}
f_{k}(\tau) = \frac{{\mathcal N_{\nu}}}{\sqrt{2 \, k\, c_{\mathrm{s}}}}\,\sqrt{k\, c_{\mathrm{s}} y} \, H^{(2)}_{\nu}
(k\, c_{\mathrm{s}} \, y), \qquad N_{\nu} = \sqrt{\frac{\pi}{2}} e^{- i \pi(2 \nu +1)/4},
\end{equation}
where, as usual, we shall focus on the case $c_{\mathrm{s}} = \sqrt{w}$.
If the initial conditions are not quantum mechanical but rather thermal, then the initial 
state will contain thermal phonons, i.e. 
\begin{equation}
\langle \hat{a}_{\vec{k}}^{\dagger} \, \hat{a}_{\vec{k}}\rangle = \overline{n}_{k} = \frac{1}{e^{k \, c_{\mathrm{s}}/\overline{T}} -1},
\label{Q4}
\end{equation}
where $\overline{T} = a T$ denotes the comoving temperature while $T$ is the physical temperature. 
The power spectrum of the fluid phonons can be easily determined from the two point function evaluated at equal 
times:
\begin{eqnarray}
\langle \hat{q}_{v}(\vec{x},\tau) \, \hat{q}_{v}(\vec{y}, \tau) \rangle &=& \int d \ln{k} \, {\mathcal P}_{q_{v}}(k,\tau) \frac{\sin{k\, r}}{k\, r},\qquad r = |\vec{x} - \vec{y}|,
\nonumber\\
{\mathcal P}_{q_{v}}(k,\tau) &=& \frac{k^3}{2 \pi^2} \bigl| f_{k}(\tau)\bigr|^2 \, (2 \overline{n}_{k} + 1).
\label{Q5}
\end{eqnarray}
If $k c_{\mathrm{s}} \gg \overline{T}$ we have that $\overline{n}_{k} \to 0$ and the quantum mechanical initial conditions 
dominate; conversely if $k c_{\mathrm{s}} \ll \overline{T}$ the thermal initial conditions dominate against 
the quantum mechanical ones. Once the phonon spectrum is known the spectrum of curvature 
perturbations is 
\begin{equation}
{\mathcal P}_{{\mathcal R}}(k,\tau) = \frac{k^3}{2 \pi^2\, z_{pr}^2} \bigl| f_{k}(\tau)\bigr|^2 \, (2 \overline{n}_{k} + 1).
\label{Q6}
\end{equation}
From the spectrum of curvature phonons it is elementary to derive the spectrum of the metric fluctuations 
and of the Bardeen potential. In fact, from Eq. (\ref{SU1}) the expression for $B(\vec{x},\tau)$ becomes
\begin{equation}
B(\vec{x},\tau) = -\frac{{\mathcal C}_{B}(w)}{\overline{M}_{\mathrm{P}}\, a^2(\tau)} \, \int^{\tau} a(\tau') \, q_{v}(\vec{x},\tau') \, d\tau',\qquad {\mathcal C}_{B}(w) = \frac{\sqrt{3 w ( w +1)}}{2}.
\label{Q7}
\end{equation}
Using Eq. (\ref{Q7})  and recalling that $\Psi(\vec{x},\tau) = - {\mathcal H} \,B(\vec{x},\tau)$ the mode expansion 
for  $\hat{\Psi}(\vec{x},\tau)$ is:
\begin{eqnarray}
\hat{\Psi}(\vec{x},\tau) &=& - \frac{{\mathcal H} {\mathcal C}_{B}(w)}{\overline{M}_{\mathrm{P}} \,a^2(\tau) \, (2\pi)^{3/2}} \int\,
d^3k\, \biggl[ \hat{a}_{\vec{k}} \, g_{k}(\tau) e^{- i \vec{k}\cdot\vec{x}} +  \hat{a}_{\vec{k}}^{\dagger} \, g_{k}^{*}(\tau) e^{ i \vec{k}\cdot\vec{x}} \biggr],
\label{Q8}\\
g_{k}(\tau) &=& \int^{\tau} a(\tau') \,\, f_{k}(\tau')\,\, d\tau'.
\label{Q9}
\end{eqnarray}
 The two-point function and the related power spectrum are simply 
 \begin{eqnarray}
\langle \hat{\Psi}(\vec{x},\tau) \, \hat{\Psi}(\vec{y}, \tau) \rangle &=& \int d \ln{k} \, {\mathcal P}_{\Psi}(k,\tau) \frac{\sin{k\, r}}{k\, r},
\nonumber\\
{\mathcal P}_{\Psi}(k,\tau) &=& \frac{k^3}{2 \pi^2} \frac{{\mathcal H}^2 \, {\mathcal C}_{B}^2(w)}{\overline{M}_{\mathrm{P}}^2 a^4(\tau)}\,\bigl| g_{k}(\tau)\bigr|^2 \, (2 \overline{n}_{k} + 1).
\label{Q10}
\end{eqnarray}
Within the same logic it is straightforward to derive the power spectrum of the density contrast. From Eqs. (\ref{SU2}) and (\ref{SU3}) after some algebra it 
can be shown that 
\begin{equation}
\delta_{\mathrm{t}}(\vec{x},\tau) =  \frac{2 \, a^{ 3 c_{\mathrm{s}}^2}}{ 3 \, {\mathcal H} \, c_{\mathrm{s}}^2} \biggl(\frac{{\mathcal H}^2 
- {\mathcal H}'}{{\mathcal H}^2} \biggr) \, \frac{\partial}{\partial\tau} \biggl[ \frac{q_{v}}{z_{pr} a^{ 3 c_{\mathrm{s}}^2}} \biggr].
\label{Q11}
\end{equation}
Finally, using Eq. (\ref{Q11}) the power spectrum of the density contrast turns out to be 
\begin{eqnarray} 
{\mathcal P}_{\delta}(k,\tau) &=& \frac{k^3}{2 \, \pi^2\,\overline{M}_{\mathrm{P}}^2} \, C_{\delta}^2(w) \bigl|\,h_{k}(\tau)\bigr|^2\, (2 \overline{n}_{k} +1),\qquad 
{\mathcal C}_{\delta}(w) = \sqrt{\frac{3(w+1)}{w}},
\nonumber\\
h_{k}(\tau) &=& \frac{a^{3 c_{\mathrm{s}}^2}}{{\mathcal H}} \, \frac{\partial}{\partial \tau} \biggl[ \frac{f_{k}(\tau)}{a^{1 + 3 c_{\mathrm{s}}^2}} \biggr].
\label{Q12}
\end{eqnarray}

Let us now suppose that the initial fluid phase is dominated by thermal phonons. In this rather 
realistic situation $w= 1/3$ and $c_{\mathrm{s}} = \sqrt{w}$. From Eqs. (\ref{Q5}) and (\ref{Q6}) 
the spectrum of curvature perturbations can be recast in the following form:
\begin{equation}
{\mathcal P}_{{\mathcal R}}(k,\,k_{H},\,T) = \frac{N_{eff}}{1440\, \sqrt{3}}\, \biggl(\frac{k}{k_{H}}\biggr)^2 \,
\biggl(\frac{T}{\overline{M}_{\mathrm{P}}}\biggr)^4 \coth{\biggl[\sqrt{\frac{\pi^2 \, N_{eff}}{360}}\, \biggl(\frac{k}{k_{H}}\biggr) \,
\biggl(\frac{T}{\overline{M}_{\mathrm{P}}}\biggr)\biggr]};
\label{Q13}
\end{equation}
where $k_{H} = a H = {\mathcal H}$; in Eq. (\ref{Q13}) the physical temperature $T$ is related to the Hubble rate as $ H^2 \overline{M}_{\mathrm{P}}^2 =  N_{eff}\, 
\pi ^2 T^{4}/30$ where $N_{eff}$ denotes the effective  number of relativistic degrees of freedom. In the limit 
$(k/k_{H}) (T/\overline{M}_{\mathrm{P}}) <1$, Eq. (\ref{Q13}) becomes:
\begin{equation}
{\mathcal P}_{{\mathcal R}}(k,\,k_{H},\,T) = \frac{\sqrt{N_{eff}}}{24 \,\sqrt{30}\, \pi} \biggl(\frac{k}{k_{H}}\biggr) \biggl(\frac{T}{\overline{M}_{\mathrm{P}}}\biggr)^{3}.
\label{Q14}
\end{equation}
When the wavenumber is of the order of the particle horizon during the protoinflationary phase the amplitude 
of the curvature phonons is solely controlled by the temperature which must not exceed the Planck 
temperature. When $k \ll k_{H}$ the power spectrum is further suppressed. Since the transition to the 
fully developed inflationary phase occurs for $\tau\simeq \tau_{*}\sim 1/(a_{*} H_{*})$ the spectrum 
computed from Eqs. (\ref{Q13}) and (\ref{Q14}) stops being valid for $k \sim k_{*} = 1/\tau_{*}$. This means that 
$(k/k_{*})<1$ implying $k/k_{H} < |\tau/\tau_{*}|$ which simply tells 
that the initial conditions during the protoinflationary phase must be set not too early or, equivalently, not too close to the Planck curvature scale. 

Similar considerations apply in the discussion the spectrum of the metric perturbations and of the density contrast. From Eqs. (\ref{Q7})--(\ref{Q10}) we arrive at the following explicit expression:
\begin{equation}
{\mathcal P}_{\Psi}(k,\, k_{H},\,T) = \frac{\sqrt{3}\, N_{eff}}{360} \,\biggl(\frac{T}{\overline{M}_{\mathrm{P}}}\biggr)^4 
\,  \biggl(\frac{3 k_{H}^2 + k^2 }{k^2} \biggr) \coth{\biggl[\sqrt{\frac{\pi^2 \, N_{eff}}{360}}\, \biggl(\frac{k}{k_{H}}\biggr) \,
\biggl(\frac{T}{\overline{M}_{\mathrm{P}}}\biggr)\biggr]}.
\label{Q15}
\end{equation}
Unlike ${\mathcal P}_{{\mathcal R}}(k,\,k_{H},\,T)$, ${\mathcal P}_{\Psi}(k,\,k_{H},\,T)$ is always a sharply decreasing function for $k > k_{H}$. 
From Eq. (\ref{Q12}) the spectrum of the density contrast is:
\begin{equation} 
{\mathcal P}_{\delta}(k,\tau) = \frac{3}{ \pi^2} \biggl(\frac{k}{k_{*}}\biggr)^4 \, \biggl(\frac{H_{*}}{\overline{M}_{\mathrm{P}}}\biggr)^2 \,\, \biggl(\frac{k^2 + 12\, k_{H}^2}{k^2}\biggr) \, 
\coth{\biggl[\sqrt{\frac{\pi^2 \, N_{eff}}{360}}\, \biggl(\frac{k}{k_{H}}\biggr) \,\biggl(\frac{T}{\overline{M}_{\mathrm{P}}}\biggr)\biggr]},
\label{Q16}
\end{equation}
where $k_{*} = a_{*} H_{*}$ and, as before, $k_{H} = a H$. From  Eq. (\ref{Q16}) it can be argued that 
as long as $k < k_{*}$ and $H_{*} < \overline{M}_{\mathrm{P}}$ the modes inside the Hubble radius (i.e. $k \gg k_{H}$) do not jeopardize the validity of the perturbative expansion.During the protoinflationary phase the Hubble rate sharply increases towards the singularity and, in this situation, it can happen that large fluctuations arise for typical scales larger than the Hubble radius.  This effect is caused by the nearness of the singularity and he lower limit in the time coordinate should be fixed by enforcing the validity of the perturbative expansion.  

\subsection{Suppression of density contrast and metric fluctuations}
During the slow-roll phase and and in the limit $z_{pr} \to 0$ the analog of Eqs. (\ref{SU1})--(\ref{SU3}) can be written as
\begin{eqnarray}
&& \dot{B} + 2 H\,B - \frac{\epsilon \,{\mathcal R}}{a} =0,
\label{SU1a}\\
&& \dot{{\mathcal R}} = \frac{\nabla^2 B}{a\, \epsilon} ,
\label{SU2a}\\
&& \delta_{\mathrm{t}} = -\frac{2}{3} \frac{\nabla^2 B}{H a} + 2 \epsilon {\mathcal R}, \qquad \delta_{\mathrm{t}} = \frac{\delta\rho_{\varphi}}{\rho_{\varphi}},
\label{SU3a}
\end{eqnarray}
where  $\delta_{\mathrm{t}}$ is now dominated by the fluctuations of the inflaton. 
We shall assume, for instance, the validity of the 
solution (\ref{EX6}) with $\beta \gg1 $. This requirement is even too restrictive since 
the results discussed hereunder are valid in the slow-roll approximation, i.e. when both slow-roll parameters\footnote{The slow-roll parameter $\epsilon$ must not be confused with the parameter of the gauge transformation introduced in Eq. 
(\ref{EB}). These two variables never appear together either in the preceding or in the following discussion so that no confusion is possible.} 
\begin{equation}
\epsilon = - \frac{\dot{H}}{H^2} = \frac{\overline{M}_{\mathrm{P}}^2}{2} \biggl(\frac{V_{,\varphi}}{V}\biggr)^2 , \qquad \eta = \frac{\ddot{\varphi}}{H \dot{\varphi}} = \epsilon - \overline{\eta}, \qquad \overline{\eta} =  \overline{M}_{\mathrm{P}}^2 
\frac{V_{,\varphi\varphi}}{V}
\label{srpar}
\end{equation}
are much smaller than $1$ but not necessarily constant. If ${\mathcal R}$ is continuous across the transition Eq. (\ref{SU1a}) implies that the expression for $B(\vec{x}, t)$ becomes:
\begin{equation}
B(\vec{x},t) = \frac{{\mathcal R}(\vec{x},\tau)}{H \, a} - \frac{1}{a^2} \int^{t} {\mathcal R}(\vec{x},t')\, dt' 
- \frac{1}{a^2(t)} \int^{t} \frac{\dot{{\mathcal R}}(\vec{x},t') \ a(t')}{H(t')} \, d\,t'.
\label{SU4}
\end{equation}
Conversely, the continuity of $B(\vec{x},t)$ in Eq. (\ref{SU2a}) implies $\dot{{\mathcal R}} \simeq 0$ for typical wavelengths larger than the Hubble radius. Therefore Eqs. (\ref{SU3}), (\ref{SU3a}) and (\ref{SU4}) imply
\begin{eqnarray}
B(\vec{x},t_{pr}) &=& \frac{{\mathcal R}_{*}(\vec{x})}{a_{pr}\, H_{pr}} \biggl[ 1 - \frac{H}{a} \int^{t_{pr}} a(t') \, \, dt'\biggr]  \simeq \frac{{\mathcal R}_{*}(\vec{x})}{a_{pr}\, H_{pr}} \,\biggl(\frac{5 + 3 w}{3  w +3 }\biggr),
\label{SU5a}\\
B(\vec{x},t_{inf}) &=& \frac{{\mathcal R}_{*}(\vec{x})}{a_{pr}\, H_{pr}} \biggl[ 1 - \frac{H}{a} \int^{t_{inf}} a(t') \, \, dt'\biggr]  \simeq\frac{{\mathcal R}_{*}(\vec{x})}{a_{inf}\, H_{inf}}\, \biggl( \frac{\epsilon}{\epsilon +1}\biggr),
\label{SU5}
\end{eqnarray}
where $t_{pr}$ and $t_{inf}$ are, respectively,  the values of the cosmic time coordinate during the protoinflationary phase and during the slow-roll phase when 
$\epsilon \ll 1$ and $\eta\ll 1$. 
Recalling that 
$\Psi(\vec{x}, t)= - (a\, H) \, B(\vec{x},t)$, Eqs. (\ref{SU5a}) and (\ref{SU5}) show that 
\begin{equation}
\Psi(\vec{x}, t_{inf}) = \frac{3(w +1)}{5 + 3 w}\, \biggl(\frac{\epsilon}{\epsilon + 1}\biggr)\,  \, \Psi(\vec{x},t_{pr}).
\label{SU8}
\end{equation}
Since during the protoinflationary phase the evolution is by definition decelerated, $0\leq w \leq 1$. Conversely 
in the inflationary regime $\epsilon \ll 1$ which demonstrates the suppression of the metric fluctuation. 
The same logic leads to following relation valid in the case of the density contrasts:
\begin{equation}
\delta_{\mathrm{t}}(\vec{x}, t_{inf})= \frac{ 3 \,\epsilon\, (w + 1)}{2}\, \delta_{\mathrm{t}}(\vec{x}, t_{pr}).
\label{SU9}
\end{equation}
Once more, since $3 (w +1)/2 \simeq {\mathcal O}(1)$ and $\epsilon \ll 1$ we also have that  
$\delta_{\mathrm{t}}(\vec{x}, t_{inf}) \ll \delta_{\mathrm{t}}(\vec{x}, t_{pr})$ 
demonstrating the suppression of the density contrast across the protoinflationary boundary. Note that the results reported here hold for a slow-roll parameter $\epsilon$ which is not necessarily constant. 
\renewcommand{\theequation}{5.\arabic{equation}}
\setcounter{equation}{0}
\section{Enhanced protoinflationary inhomogeneities}
\label{sec5}
\subsection{Monotonicity properties}
The dynamics of the protoinflationary transition depends on the behaviour of the pump fields $z_{pr}(t)$ and $z_{\varphi}(t)$. In the sudden treatment of the transition 
discussed in section \ref{sec4} the pump fields $z_{pr}(t)$ and $z_{\varphi}(t)$ 
evolve monotonically in cosmic time and 
the corresponding rates are positive definite (i.e. $\dot{z}_{pr}/z_{pr} >0$ and $\dot{z}_{\varphi}/z_{\varphi} >0$). Furthermore the evolution 
of $z_{pr}(t)$ is  decelerated (i.e. $\ddot{z}_{pr}(t) <0$).
These requirements are satisfied in section \ref{sec4} where  $z_{pr}(t) \propto a(t)$, the evolution is decelerated as long as the perfect barotropic fluid dominates (i.e. $\dot{z}_{pr}(t) > 0$ and $\ddot{z}_{pr}(t) < 0$); furthermore, recalling Eq. (\ref{srpar}), during the slow-roll phase
\begin{equation}
\frac{\dot{z_{\varphi}}}{z_{\varphi}}= H\biggl[ 1 + \frac{\ddot{\varphi}}{H \dot{\varphi}} - \frac{\dot{H}}{H^2} \biggr] 
= H ( 1 + \eta + \epsilon) >0, 
\label{trans1}
\end{equation}
since, by definition of slow-roll, $\epsilon \ll 1$ and $\eta \ll 1$ with $\epsilon > 0$. In the particular case of exponential potentials the slow-roll parameters are constant.
The monotonicity requirements define the conventional protoinflationary dynamics and they seem to be naively sufficient to implement a successful transition. Are they at all necessary? Is it possible to implement the transition 
in a different way without relying on the monotonicity of the pump fields?  
If this is the case, which are the consequences for the evolution of curvature perturbations?

If the transition to the  slow-roll phase is not sudden
the behaviour of $z_{pr}(t)$ and $z_{\varphi}(t)$ is not necessarily monotonic. 
Whenever $z_{pr}(t)$ and $z_{\varphi}(t)$ do not evolve monotonically they can develop either a global maximum or a global minimum. There can be even more cumbersome evolutions where a finite number of maxima and minima arise. These situations can be discussed after having addressed the basic case where either  $\dot{z}_{pr}(t)$ or $\dot{z}_{\varphi}(t)$ (or both) vanish 
for a finite value of the cosmic time coordinate $t$. The absence of monotonic behaviour is dynamically realized in different ways by appropriately modifying the relative weight of the inflationary and fluid components in the total energy-momentum tensor of the system.  A class of analytic solutions exhibiting  non-monotonic behaviour  for the pump fields 
can be obtained by solving Eqs. (\ref{FL1})--(\ref{FL4}) whose 
explicit form, in the cosmic time coordinate, is: 
\begin{eqnarray}
&& 3 H^2\, \overline{M}_{\mathrm{P}}^2 = \frac{\dot{\varphi}^2}{2} + V(\varphi) + \rho_{pr},
\label{FL1a}\\
&& 2\,\dot{H}\, \overline{M}_{\mathrm{P}}^2  = - \dot{\varphi}^2 - (p_{pr} + \rho_{pr}),
\label{FL2a}\\
&& \ddot{\varphi} + 3 H \dot{\varphi} + \frac{\partial V}{\partial \varphi} =0,
\label{FL3a}\\
&& \dot{\rho}_{pr} + 3 H ( p_{pr} + \rho_{pr}) =0.
\label{FL4a}
\end{eqnarray}
We shall be interested in solutions where $z_{pr}(t)$ and $z_{\varphi}(t)$ are not monotonic but satisfy the correct boundary conditions typical of a protoinflationary dynamics. Equation (\ref{FL2a}) can be written in terms of $z_{\varphi}(t)$ and $z_{pr}(t)$ and using the definition of $\epsilon(t)$ given in Eq. (\ref{srpar})
\begin{equation}
a^2 \overline{M}_{\mathrm{P}}^2 \epsilon(t) = z^2_{\varphi}(t) + c_{\mathrm{s}}^2 z^2_{pr}(t),\qquad \epsilon(t) = - \frac{\dot{H}}{H^2}. 
\label{SOL0}
\end{equation}
\subsection{Exact solutions}
By requiring that $z_{pr}(t)$ and $z_{\varphi}(t)$ are proportional it is possible to obtain a suitable ansatz for the 
solution of the whole system subjected to the requirement that the evolution of $z_{\varphi}(t)$ and   $z_{pr}(t)$ is not monotonic. In the case of perfect barotropic fluid with constant sound speed the full solution of Eqs. (\ref{FL1})--(\ref{FL2}) and (\ref{FL3})--(\ref{FL4})  can be expressed in terms of the scale factor and of the 
inflaton field:
\begin{eqnarray}
&& a(t) = a_{*} \biggl[ \sinh{(\beta \, H_{*}\, t)}\biggr]^{1/\beta}, \qquad \beta = \frac{3 ( w + 1)}{2},
\label{SOL1}\\
&& \varphi(t) = \varphi_{0} \pm \sqrt{\frac{2}{\beta}} \overline{M}_{\mathrm{P}} \sqrt{1 - \Omega_{*}} \ln{\biggl[\tanh{\biggl( \frac{\beta H_{*} t}{2}\biggr)}\biggr]}, 
\label{SOL2}
\end{eqnarray}
where the parameter $\Omega_{*}$ and the protoinflationary energy density are defined, respectively,  as
\begin{equation}
\Omega_{*}= \frac{\rho_{*}}{3 H_{*}^2 \overline{M}_{\mathrm{P}}^2},\qquad \rho(t)= \rho_{*} \biggl(\frac{a_{*}}{a}\biggr)^{ 3 ( w +1)}.
\label{SOL4a}
\end{equation}
Finally the inflaton potential is 
\begin{equation} 
V(\varphi) = 3 H_{*}^2 \overline{M}_{\mathrm{P}}^2 + 
\frac{3}{2} (1 - w) H_{*}^2 \overline{M}_{\mathrm{P}}^2 (1 - \Omega_{*}) \sinh^2{\biggl[\sqrt{\frac{\beta}{2}} \frac{(\varphi - \varphi_{0})}{(1 - \Omega_{*}) \overline{M}_{\mathrm{P}}}\biggr]}.
\label{SOL4}
\end{equation}
Equations (\ref{SOL1})--(\ref{SOL4}) 
solve Eqs. (\ref{FL1a})--(\ref{FL4a}) in the case of a constant 
barotropic index. Furthermore, as anticipated, the solution 
satisfies the boundary conditions characterizing the protoinflationary 
transition. In particular for $\beta H_{*} t < 1$ the solution is decelerated and from Eq. (\ref{SOL1}) we have
\begin{equation}
a(t) \simeq a_{*} ( \beta H_{*} t)^{1/\beta}, \qquad H_{i} = \frac{1}{\beta t_{i}} =\frac{2}{3 ( w+ 1) t_{i}}.
\label{SOL5}
\end{equation}
In the opposite limit (i.e. $\beta H_{*} t \gg 1$) the solution is accelerated with $H(t) \simeq H_{*}$:
\begin{equation}
H(t) = \frac{H_{*}}{\tanh{(\beta H_{*} t)}}, \qquad \dot{H} = - \frac{\beta H_{*}^2}{\sinh^2{(\beta H_{*} t)}}.
\label{SOL6}
\end{equation}
The parameter $\Omega_{*} = \rho_{*}/( 3 H_{*}^2 \overline{M}_{\mathrm{P}}^2) < 1$ measures 
the amount of radiation at the onset of inflation. From Eq. (\ref{SOL5}),  $H_{*}$ is not the initial curvature scale but rather the curvature scale at the onset of the inflationary evolution. 
The conditions $\ddot{a} >0$ and $\dot{a}>0$ imply, from Eq. (\ref{SOL1}),
\begin{equation}
\ddot{a} = \frac{a_{*} H_{*}^2}{2} \bigl[\sinh{(H_{*} t \beta)}\bigr]^{1/\beta -2} 
\biggl\{ 1 - 2 \beta + \cosh{[2 H_{*} \beta t]}\biggr\}>0.
\label{cond1}
\end{equation}
From Eq. (\ref{cond1}) $\ddot{a} >0$ iff 
$\cosh^2(\beta H_{*} t) > (\beta-1)$. The beginning of the inflationary phase $t_{x}$
is then given by
\begin{equation}
H_{*} t_{x} = \frac{1}{\beta} \ln{[\sqrt{\beta} + \sqrt{\beta -1}]}.
\label{cond2}
\end{equation}
This requirement shows that $t_{x}$ ranges between $0.54\, H_{*}^{-1}$ in the case $w=1/3$ and and 
$0.49\, H_{*}^{-1}$ in the case $w=1$. For numerical purposes related to the evolution of the inhomogeneities (see 
the discussion hereunder) it is practical to use $H_{*} \, t$ as new time coordinate. In fact $H_{*} t$ approximately coincides with the natural logarithm of the total number of inflationary efolds. Hence $H_{*} t_{\mathrm{max}} \simeq {\mathcal O}(\ln{N_{\mathrm{tot}}})$ where we can take, for illustrative purposes $N_{\mathrm{tot}}$ between $70$ and $100$. Equations (\ref{srpar}) and (\ref{SOL1})--(\ref{SOL2}) imply that $\epsilon(t)$ gets progressively much smaller than $1$ for $\beta \,H_{*} \, t \gg 1 $
while $\overline{\eta}(t)$ and $\eta(t)$ are nearly constant in the same limit:
\begin{equation}
\epsilon(t)= \frac{\beta}{\cosh^2{(\beta \, H_{*} \,t)}}, \qquad \eta(t) = - \beta, \qquad \overline{\eta}(t) = \beta \frac{[1 + \cosh^2{(\beta \, H_{*} \,t)}]}{\cosh^2{(\beta \, H_{*} \,t)}}.
\end{equation}
In conclusion the solution of Eqs. (\ref{SOL1})--(\ref{SOL4}) satisfies 
the physical properties of a protoinflationary regime and can be used to describe 
the transition to from a decelerated epoch to an accelerated phase where 
the slow roll condition is verified for $\epsilon$ but not for $\eta$. Notice, finally, that when $\Omega_{*} \ll 1$ 
the potential $V(\varphi)$ of Eq. (\ref{SOL4}) can always be expanded in powers of $\Omega_{*}$ as 
\begin{equation}
V(\varphi) = V_{0}(\varphi) + V_{1}(\varphi) \Omega_{*} + V_{2}(\varphi) \Omega_{*}^2 + ...
\end{equation}
where the functions $V_{0}(\varphi)$, $V_{1}(\varphi)$,  $V_{2}(\varphi)$ are uniquely 
fixed from the exact form of the potential. 

\begin{figure}[!ht]
\centering
\includegraphics[height=6cm]{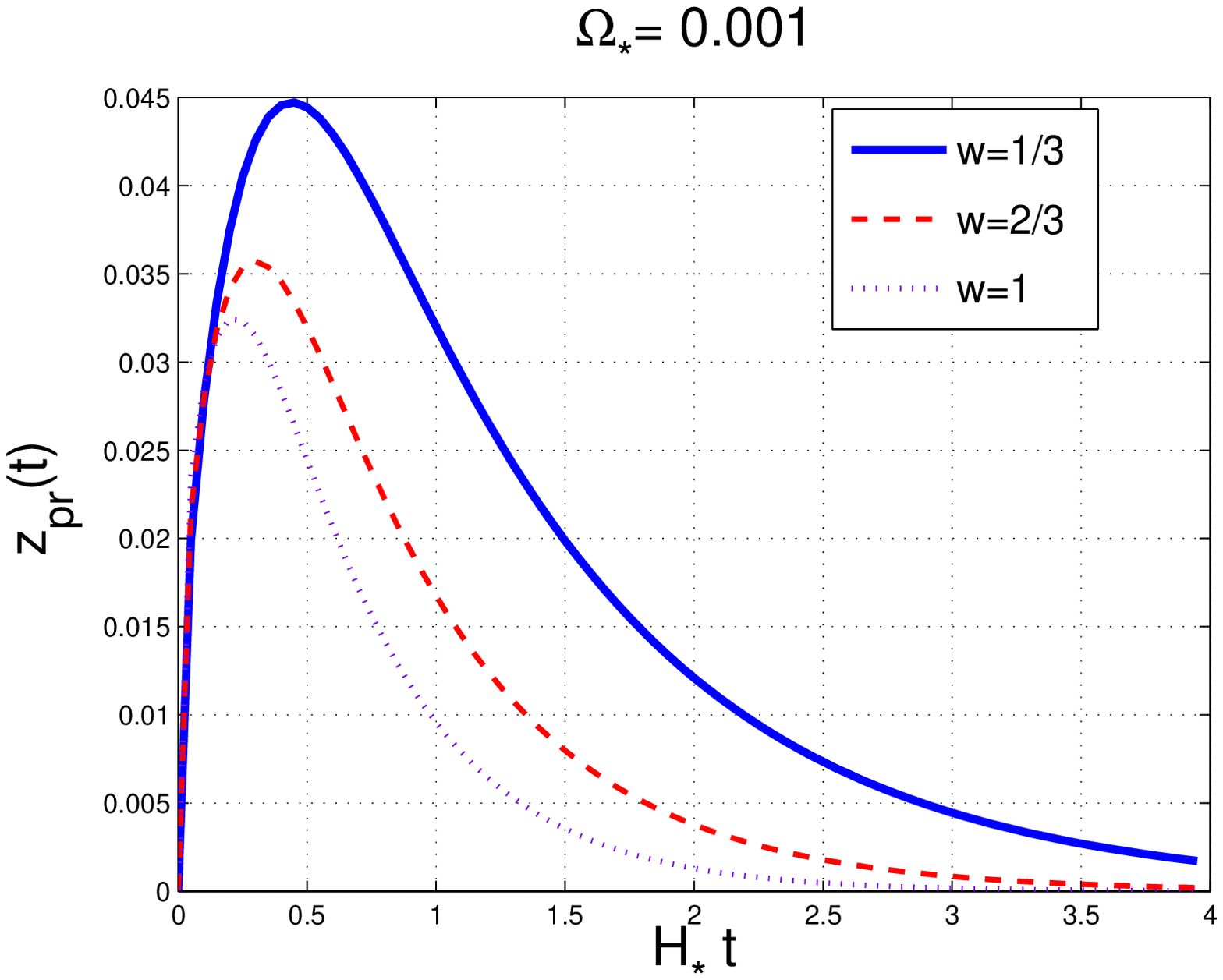}
\includegraphics[height=6cm]{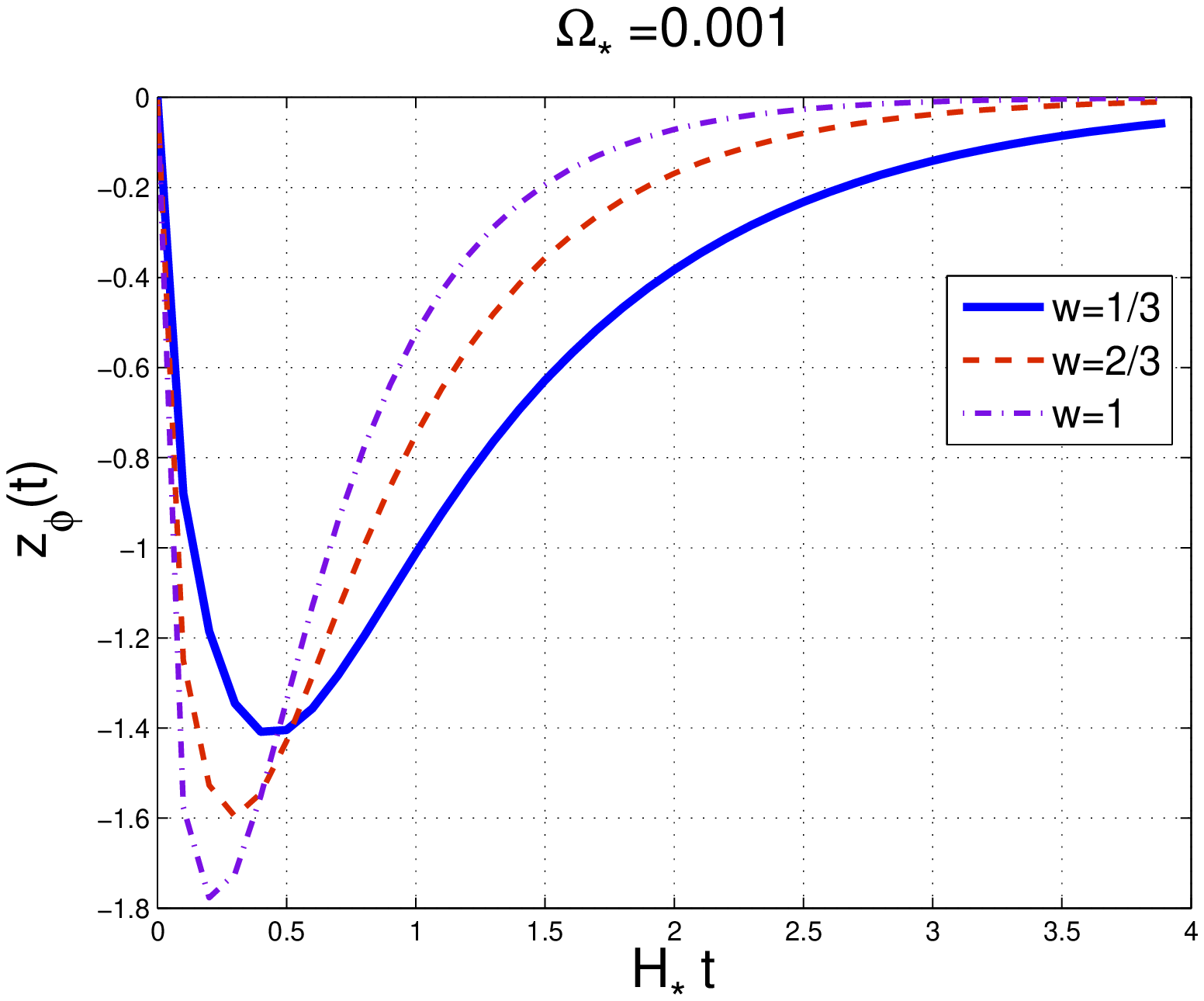}
\caption[a]{The evolution of $z_{pr}(t)$ and of $z_{\varphi}(t)$ is illustrated in the case of the exact solution 
reported in Eqs. (\ref{SOL1}) and (\ref{SOL2}).}
\label{figure1}      
\end{figure}

The class of solutions reported in Eqs. (\ref{SOL1})--(\ref{SOL4}) can be used to determine the explicit forms of $z_{pr}(t)$ and of $z_{\varphi}(t)$:
\begin{equation}
z_{pr}(t) = \sqrt{\frac{3( 1+ w)}{w}} \sqrt{\Omega_{*}} \frac{a(t)}{\cosh{(\beta H_{*} t)}}\, \, \overline{M}_{\mathrm{P}},\qquad \frac{z_{pr}(t)}{z_{\varphi}(t)} = - \frac{1}{\sqrt{w}} \frac{\sqrt{\Omega_{*}}}{\sqrt{1 - \Omega_{*}}}.
\label{zzz}
\end{equation}
Equations (\ref{zzz}) are illustrated in Fig. \ref{figure1} where the non-monotonic behaviour of the pump fields is evident. In Fig. \ref{figure1} the scale is linear on both axes and the 
three curves correspond to different choices of the barotropic indices. Since the plots of Fig. \ref{figure1}
are presented in terms of $H_{*} t$,  they hold for any value 
of $H_{*}$. In spite of this the value of $H_{*}$ must be assigned in the numerical integration 
of the evolution of the inhomogeneities (see the discussion reported hereunder) since t controls, ultimately, the 
absolute normalization of the power spectra. As Eq. (\ref{zzz}) shows, 
the value of $\Omega_{*}$ determines the relative magnitude of $z_{pr}$ and $z_{\varphi}$.
The difference in the overall sign of $z_{pr}(t)$ and $z_{\varphi}(t)$ comes from the sign of $\dot{\varphi}$ which has been taken to be negative in Eq. (\ref{SOL2}). A flip in the sign of $\dot{\varphi}$ entails a flip in the sign 
of $z_{\varphi}(t)$.  
\subsection{Numerical integrations}
The system of Eqs. (\ref{NMM1}) and (\ref{NMM2}) will now be integrated in Fourier space and in the cosmic time parametrization which is preferable since the 
class of exact solutions reported in Eqs. (\ref{SOL1})--(\ref{SOL4}) has a simpler 
expression in cosmic time. Equations (\ref{NMM1}) and (\ref{NMM2}) can be written in the form of a plane autonomous system:
\begin{eqnarray}
\dot{q}_{v} &=& p_{v}, 
\label{NN0}\\
\dot{q}_{\chi} &=& p_{\chi},
\label{NN1}\\
\dot{p}_{v} &=& - H p_{v} + A_{v\,v}(k,t) q_{v} + B_{v\,\chi}(t) q_{\chi} + C_{v\,\chi}(t) p_{\chi},
\label{NN2}\\
\dot{p}_{\chi} &=&  - H p_{\chi} + \overline{A}_{\chi\,\chi}(k,t) q_{\chi} + \overline{B}_{\chi\,v}(t) q_{v} + \overline{C}_{\chi\,v}(t) p_{v}.
\label{NN3}
\end{eqnarray}
Equations (\ref{NN0})--(\ref{NN3}) hold in Fourier space, i.e. $q_{v}\equiv q_{v}(k,t)$, $q_{\chi}\equiv q_{\chi}(k,t)$
and similarly for $p_{v}$ and $p_{\chi}$. For notational convenience and to avoid the proliferation 
of subscripts the reference to the modulus of the wavenumber $k$ has been omitted unless 
strictly necessary. 
The coefficients  $A_{v\,v}(k,t)$,  $B_{v\,\chi}(t)$ and  $C_{v\,\chi}(t)$ are given, respectively, by:
\begin{eqnarray}
A_{v\,v}(k,t)&=& \frac{3 w -1}{2} \dot{H} + \frac{(3 w-1) (3 w+1)}{4} H^2 - \frac{c_{\mathrm{s}}^2 k^2}{a^2} 
\nonumber\\
&+& \frac{p + \rho}{4 H^2} \biggl[(w-1) \dot{\varphi}^2 - 2 \biggl( 3( w+1) H^2 + 2 \dot{H}\biggr)\biggr],
\label{NN4}\\
B_{v\,\chi}(t) &=& \frac{\sqrt{p_{pr} + \rho_{pr}}}{4 H \, \sqrt{w}} \biggl\{ \biggl(\frac{\dot{\varphi}}{H} \biggr) \biggl[(w-1) \dot{\varphi}^2 - 2 \biggl( 3( w+1) H^2 + 2 \dot{H}\biggr)\biggr]
\nonumber\\
&-& 2 (w+1) \frac{\partial V}{\partial\varphi} + 2 (1 - w) H \dot{\varphi}\biggr\},
\label{NN5}\\
C_{v\,\chi}(t)&=& - \frac{\sqrt{p_{pr} + \rho_{pr}}}{2 H\, \sqrt{w}}  (1 - w) \dot{\varphi}
\label{NN6}
\end{eqnarray}
The coefficients $\overline{A}_{\chi\,\chi}(k,t)$, $\overline{B}_{\chi\,v}(t)$ and $\overline{C}_{\chi\,v}(t)$ are instead given by:
\begin{eqnarray}
\overline{A}_{\chi\,\chi}(k,t) &=& - \frac{\partial^2 V}{\partial \varphi^2}  + (\dot{H} + 2 H^2) - \frac{k^2}{a^2}
\nonumber\\
&-& \frac{\dot{\varphi}}{4 H} \biggl[ 8 \frac{\partial V}{\partial \varphi} + \biggl(\frac{\dot{\varphi}}{H} \biggr)\biggl( 4 (\dot{H} + 3 H^2) + (p_{pr} + \rho_{pr}) \biggl(\frac{w-1}{w} \biggr) \biggr) \biggr] 
\label{NN7}\\
\overline{B}_{\chi\,v}(t)&=&  - \frac{\sqrt{p_{pr} + \rho_{pr}}}{4 H} \biggl\{ \frac{(w -1)(3 w-1)}{w} \dot{\varphi} H + 4 \frac{\partial V}{\partial\varphi} 
\nonumber\\
&+& \biggl(\frac{\dot{\varphi}}{H} \biggr)\biggl[ 4 (\dot{H} + 3 H^2) + (p_{pr} + \rho_{pr}) \biggl(\frac{w-1}{w} \biggr)\biggr]\biggr\} \,\sqrt{w}
\label{NN8}\\
\overline{C}_{\chi\,v}(t)&=& \frac{\sqrt{p_{pr} + \rho_{pr}}}{2 H} \biggl(\frac{w-1}{\sqrt{w}} \biggr) \dot{\varphi}.
\label{NN9}
\end{eqnarray}
\begin{figure}[!ht]
\centering
\includegraphics[height=6cm]{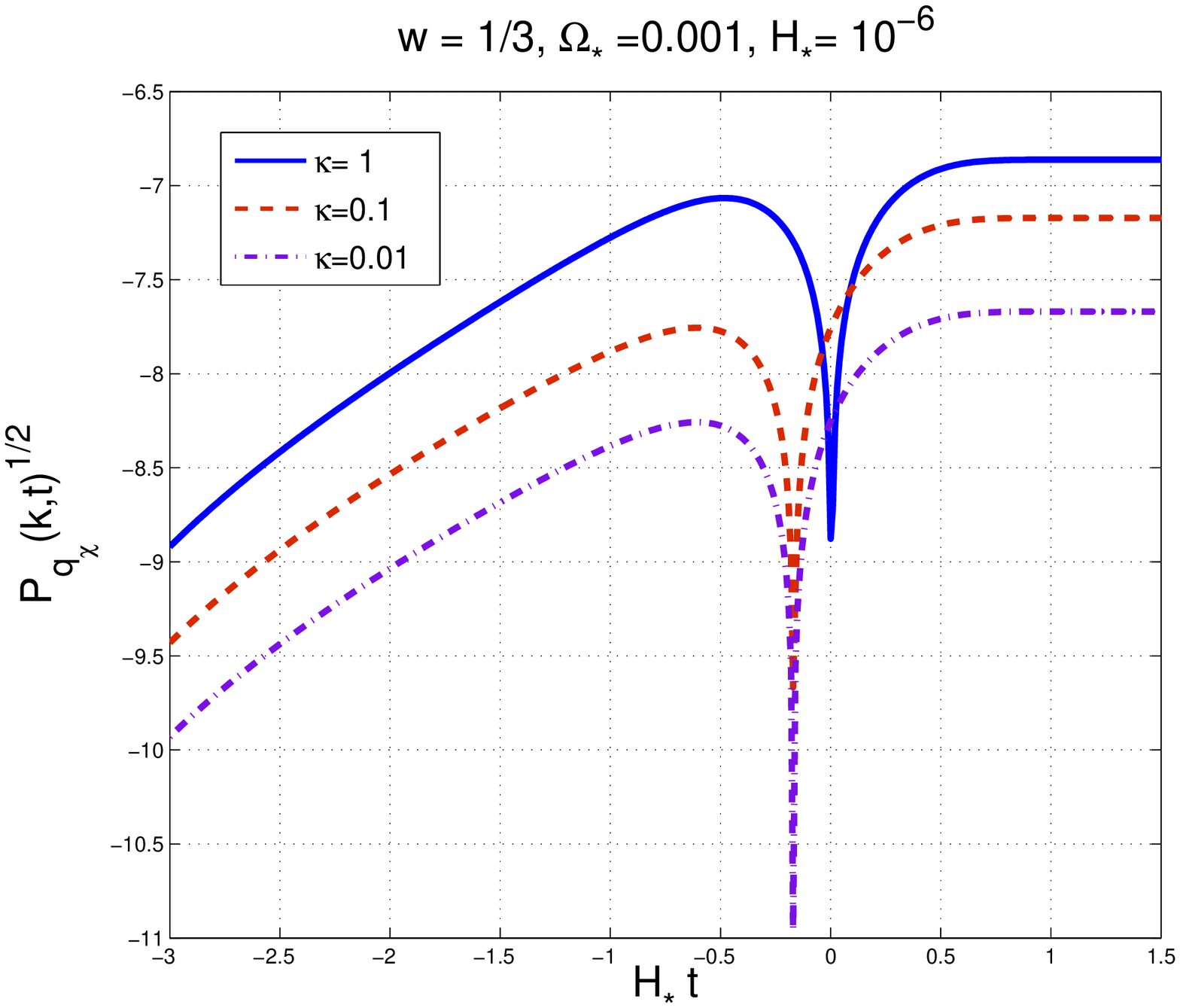}
\includegraphics[height=6cm]{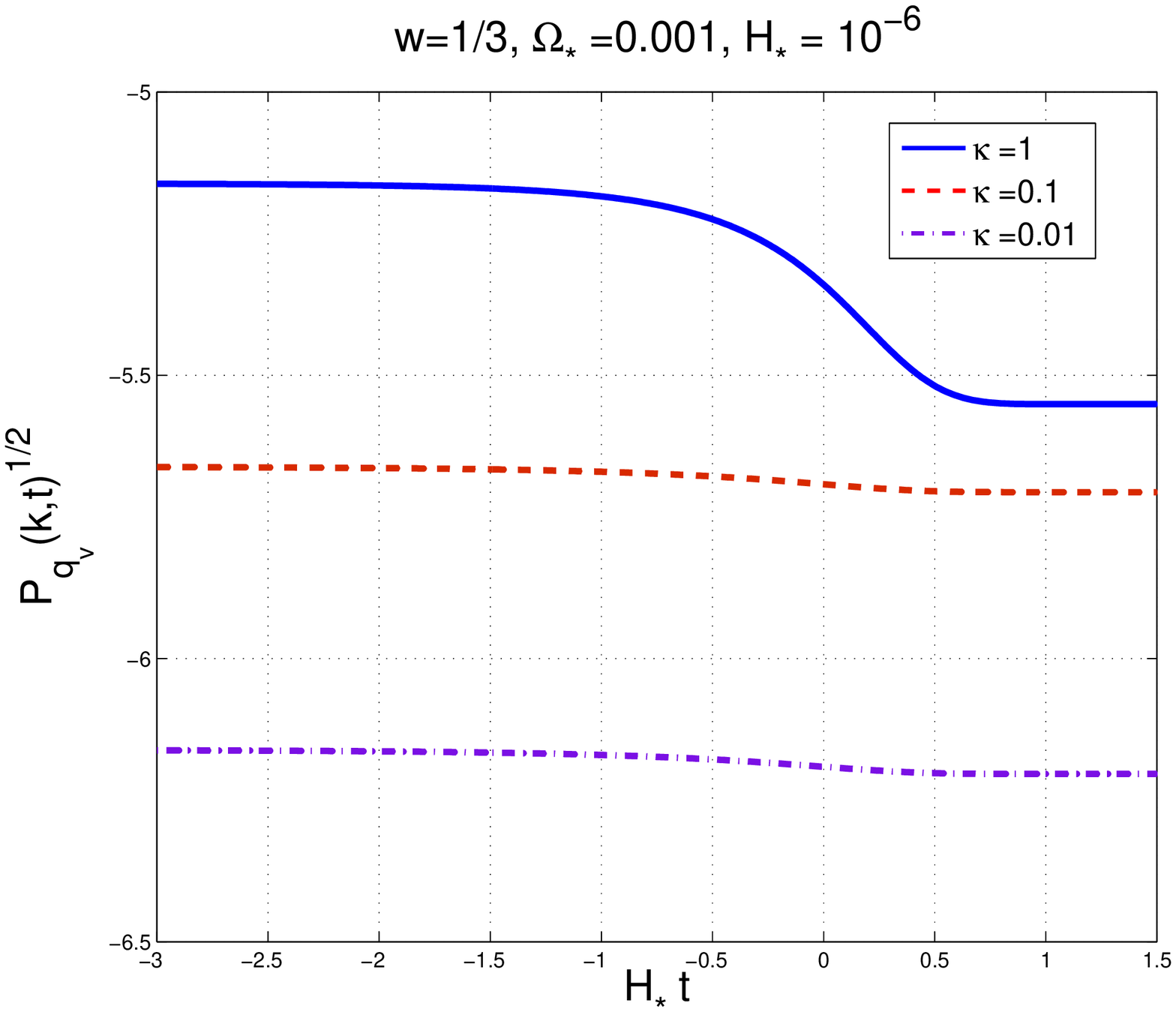}
\caption[a]{The spectrum of $q_{\chi}$ (plot at the left) and of $q_{v}$ (plot at the right). Common logarithms are 
of the corresponding quantities are used on both axes.}
\label{figure2}      
\end{figure}
Equations (\ref{NN4})--(\ref{NN6}) and Eqs. (\ref{NN7})--(\ref{NN9}) are 
related to Eqs. (\ref{A1})--(\ref{C1}) and (\ref{A2})--(\ref{C2}) modulo an overall sign
difference\footnote{Note 
that in Eqs. (\ref{NMM1}) and (\ref{NMM2}) the coefficients all appear at the left hand side in the corresponding equations. In Eqs. (\ref{NN2}) and (\ref{NN3}) the coefficients appear at the right hand side.} and the obvious algebraic changes related to the differences between the conformal and the cosmic time parametrization. Notice, furthermore, that 
the coefficients $A_{v\,v}(k,t)$ and $\overline{A}_{\chi\,\chi}(k,\tau)$ depend both on the cosmic time coordinate and on 
the wavenumber which comes from the three-dimensional Laplacian of $q_{v}$ and $q_{\chi}$.

The initial conditions for the numerical integration of Eqs. (\ref{NN0})--(\ref{NN3}) are fixed by 
requiring that, at the initial integration time, $q_{\chi}= p_{\chi} =0$ while
$q_{v}$ and $p_{v}$ are determined from the thermal phonons $\overline{n}_{k}$ during a radiation-dominated regime (i.e. $w=1/3$). The considerations 
of section \ref{sec4} and the exact solutions presented in this section
allow for different values of the barotropic indices and for a wider 
set of initial conditions. In spite of this it seems less 
confusing, for the purposes of the presentation, 
not to indulge in an excessive multiplication of examples which could 
cast shadow over the main motivation of the present analysis.
 
\begin{figure}[!ht]
\centering
\includegraphics[height=6cm]{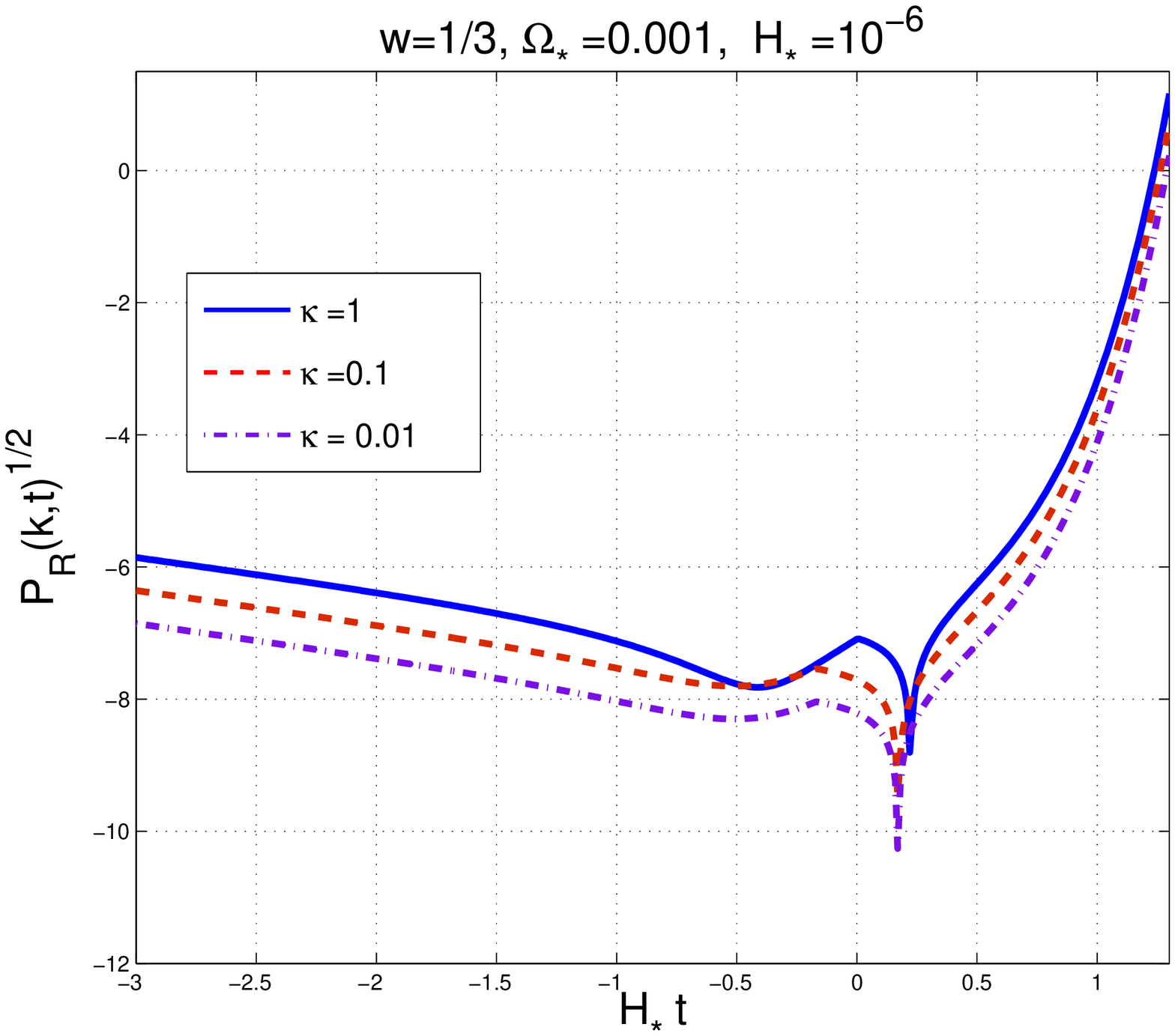}
\includegraphics[height=6cm]{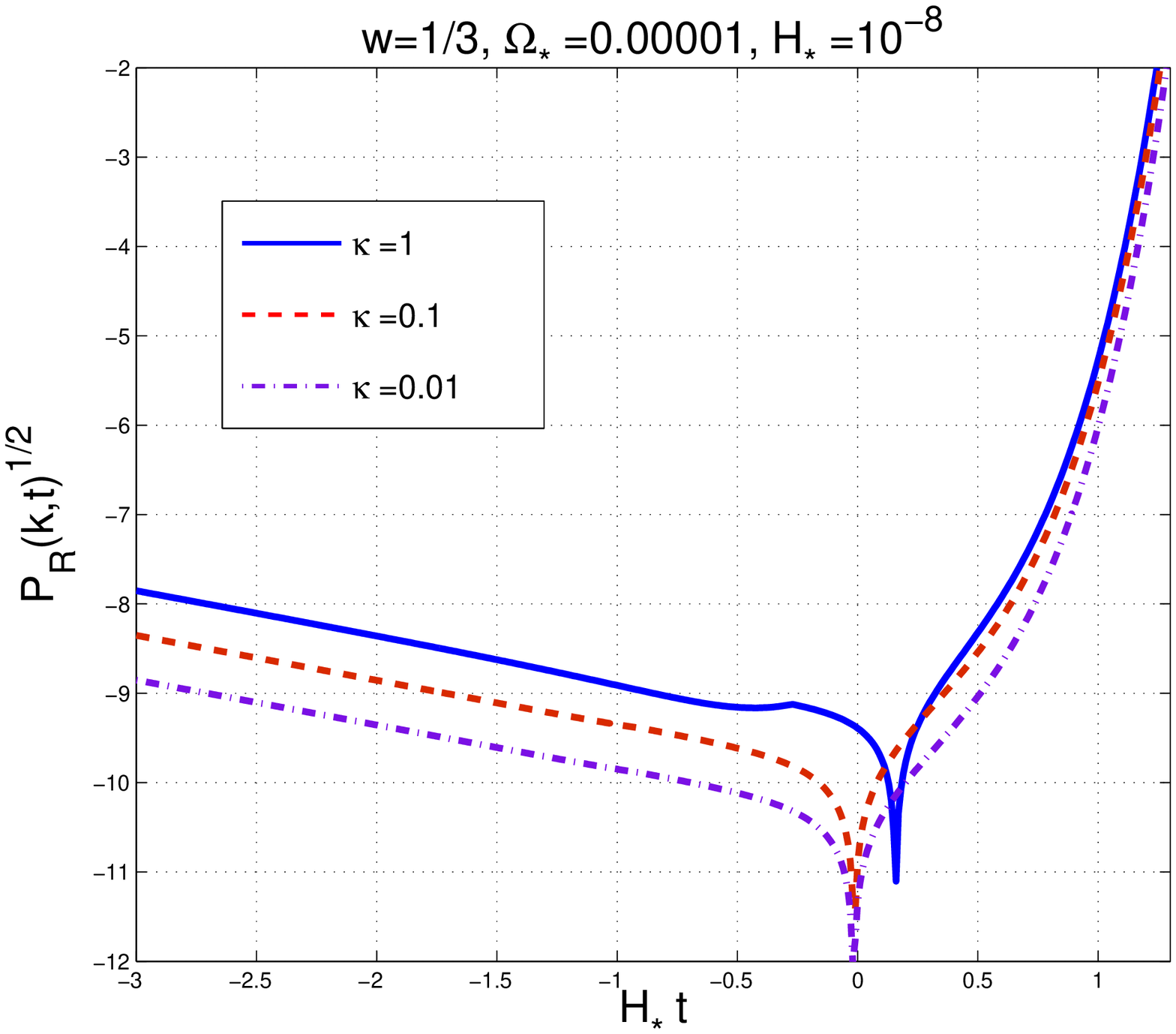}
\caption[a]{The spectra of the curvature perturbations on comoving orthogonal 
hypersurfaces for slightly different sets of fiducial parameters. As in Fig. \ref{figure2} common logarithms
of the corresponding quantity are used on both axes.}
\label{figure3}      
\end{figure}

Since the initial conditions 
are given during the protoinflationary phase (i.e. $H_{*} t \ll 1$), nearly all the relevant modes are larger than the particle horizon since $k/a(t_{i}) \ll H(t_{i})$. This aspect can be understood by looking at the sharp increase of $H(t)$ in the 
limit $t\to 0$ where the Hubble rate  dominates over $1/a(t)$ in spite of the possible 
largeness of the wavenumber. The full, dashed and dot-dashed  curves appearing
in the plots of Figs. \ref{figure2}  and \ref{figure3} 
correspond to different values of the wavenumbers $k$ which are 
expressed in units of $H_{*}$ in terms of the rescaled wavenumber  $\kappa = k/H_{*}$ (see also the legends of each plot).  The values of $H_{*}$ used in each numerical integration are reported, in natural Planckian units $\overline{M}_{\mathrm{P}} =1$, in the title of each plot 
illustrated in Figs.  \ref{figure2}  and \ref{figure3}.

As explained in sections \ref{sec3} and \ref{sec4}, once the spectra of $q_{\chi}$ and $q_{v}$ are known, all the other quantities can be computed such as the spectra of $B$ and $\phi$ or the spectra of $\Psi$ and ${\mathcal R}$. 
On both axes of Figs. \ref{figure2} and \ref{figure3} the common logarithm 
of the corresponding quantity is reported. Instead of llustrating the 
power spectrum it is more practical to compute the square root of it, as indicated 
on the vertical axes of Figs. \ref{figure2} and \ref{figure3}. During the protoinflationary 
phase (i.e. roughly $H_{*} t < 1$) the spectrum goes as $\kappa^{1/2}$ as it can be deduced from the different curves holding for decreasing values of $\kappa$. 
Recall that in Figs. \ref{figure2} and \ref{figure3}  the common logarithm of the corresponding quantity is reported on both axes.
The slope $\kappa^{1/2}$ arises since, initially, $k/( 2 T a) \ll 1$ and therefore 
${\mathcal P}_{q_{v}}^{1/2} \propto k \sqrt{T a/k}$. As $H_{*} t \gg1$ the spectrum tends to change even if 
this aspect is more visible by looking directly at the spectrum of ${\mathcal R}$ reported in Fig. \ref{figure3}.
From Fig. \ref{figure3} the spectrum of curvature perturbations exhibits a sudden growth which can even be understood on a qualitative basis. Recall, in this respect,  the general expression of ${\mathcal R}(t)$
in terms of $q_{v}$ and $q_{\chi}$ (see Eq. (\ref{rgen})). 
When the spectrum of $q_{\chi}$ and $q_{v}$ is bounded (see Fig. \ref{figure2}) ${\mathcal P}_{{\mathcal R}}$ increases when 
either $z_{pr}(t)$ or $z_{\varphi}(t)$ decrease. The slope of the spectrum goes as $\kappa^{1/2}$ for 
$H_{*} t < 1$ while it is quasi-flat for $H_{*} t > 1$. In the left plot of Fig. \ref{figure3} the parameters 
$\Omega_{*}$ and $H_{*}$ coincide with those adopted in Fig. \ref{figure2}; in the right plot of Fig. \ref{figure3} the values 
of the fiducial parameters are instead $\Omega_{*} = 10^{-4}$ and $H_{*} = 10^{-8}$. The examples discussed here are just meant to show that the lack of monotonicity of $z_{pr}(t)$ and $z_{\varphi}(t)$ 
always entails a potential decrease of the pump fields and, hence, a potential 
increase of the total curvature perturbations.
It is interesting to point out that the possibility of anomalous growth of curvature 
perturbations in single field inflationary models has been discussed some time ago but within a different perspective \cite{LLS} and not in connection with the protoinflationary dynamics.

The analysis of this section has been conducted in the case of a single protoinflationary fluid and a single inflaton; there seems to be however no obstruction for the possible generalization of this approach  
to  a finite number of perfect fluids simultaneously present together with a finite number of inflaton fields. 
Finally, the considerations of this paper just assumed the validity 
of the perturbative expansion prior to the onset of the inflationary evolution. The overall consistency of the approach, as  already stressed in section \ref{sec4}, implies that valid conclusions can only be drawn in the regions of the parameter space where the perturbative expansion is valid. At the same time there are techniques, like the gradient expansion, which can be applied in situations where the conventional perturbative expansion fails \cite{grad1,grad2,grad3}. This analysis is anyway beyond the scopes of the present paper.
\renewcommand{\theequation}{6.\arabic{equation}}
\setcounter{equation}{0}
\section{Concluding remarks}
\label{sec6}
The effects of a dynamical phase preceding 
inflation are customarily parametrized in terms of an initial number of inflaton quanta modifying the standard vacuum initial conditions for the evolution of the  large-scale inhomogeneities. In this paper the implications of a protoinflationary phase of decelerated expansion have been examined with the purpose of relaxing the 
standard lore. The present approach stipulates that, initially, the energy-momentum tensor is dominated by an irrotational fluid. After a transient regime (which may be either sharp or delayed) the inflaton potential dominates the total energy-momentum tensor and the slow-roll dynamics starts off. 
The initial conditions for the large-scale inhomogeneities  during the protoinflationary phase can be either quantum mechanical or, more realistically, thermal if the protoinflationary phase is dominated by radiation. 

Is the existence of a protoinflationary phase just equivalent to a modified set of initial conditions manually imposed on the modes of the perturbations? The results of the present analysis show that the answer this question is positive 
insofar as the protoinflationary transition is sufficiently sharp and satisfies a number of monotonicity 
requirements. Across the protoinflationary boundary it is plausible to demand that the extrinsic curvature is continuous while the stress tensor undergoes a finite discontinuity on the constant energy density hypersurface.  In this picture the transition is 
sharp and both the density contrast and the metric fluctuations are suppressed across the transition. This sudden approximation merely extends to the protoinflationary boundary the standard discussion of the various post-inflationary transitions. 

The dynamics of the inhomogeneities  is determined by the evolution of the two quasi-normal modes of the system. Depending on the time evolution of the pump fields controlling the contribution of the quasi-normal modes to the curvature perturbations, the protoinflationary transition is naturally classified in two broad classes. In the first category the pump fields are monotonic and the quasi-normal modes of the system are nearly not interacting. The second category contemplates the complementary situation where the transition is not monotonic and the pump fields have, at least, either a maximum or a minimum. In the latter case large-scale curvature perturbations can be enhanced in comparison with the results obtainable in the case of monotonic transitions. 

The large-scale modifications of the temperature and polarization power spectra are often engineered
by assuming that the onset of inflationary dynamics is only characterized by a single inflaton field
possibly with non-standard initial conditions for its inhomogeneities. The 
exact solutions obtained in this paper, as well and the numerical treatment of the corresponding large-scale 
inhomogeneities, suggest that the protoinflationary dynamics is richer than expected. 
In a more conservative perspective the obtained results show that the potential enhancement of large-scale curvature perturbations can be avoided by demanding that the evolution of the pump fields is strictly monotonic. 
It is however unclear whether or not the monotonicity of the pump fields must be considered a generic feature of the 
protoinflationary dynamics.

\section*{Acknowledgments}
It is a pleasure to thank T. Basaglia and A. Gentil-Beccot of the CERN scientific information service for their kind assistance.

\newpage
\begin{appendix}
\renewcommand{\theequation}{A.\arabic{equation}}
\setcounter{equation}{0}
\section{Explicit derivation of the quasi-normal modes}
\label{APPA}
In this appendix the details of the derivations leading to Eqs. (\ref{NMM1}) and 
(\ref{NMM2}) will be outlined.  By taking the first time derivative of Eq. (\ref{th}) the following equation arises:
\begin{equation}
u'' + {\mathcal H}' ( 1 - 3 w) u + {\mathcal H}(1 - 3 w) u' = \biggl[\frac{w}{w + 1} \delta' + \phi' \biggr],
\label{QN10}
\end{equation}
where, as already mentioned in section \ref{sec3}, the barotropic index $w$ has been taken to be constant 
so that $c_{\mathrm{s}} = \sqrt{w}$; note that $\delta = \delta\rho_{pr}/\rho_{pr}$  denotes, only in this appendix, the density contrast of the protoinflationary fluid.
Now the logic is to express the right hand side of Eq. (\ref{QN10}) solely in terms of $u$ and $\chi$. The same procedure must then be applied in the case of Eq. (\ref{chi}).  

The right hand side of Eq. (\ref{QN10}) is expressible in terms of $u$, $\chi$ 
and $\chi'$. To achieve this goal, Eqs. (\ref{HG1}) and (\ref{dr}) can be summed up term by term.
For a perfect barotropic fluid
Eq. (\ref{dr}) is given by  $\delta' =  - (1 + w) \theta_{pr}$;  by then using the expression for $\phi'$ obtainable from Eq. (\ref{sep3}) it is easy to show, after some algebra, that
\begin{eqnarray}
\phi' + \frac{w}{w+1} \delta' &=& w \nabla^2 u
+  \biggl\{ w \biggl[ \frac{4\pi G {\varphi'}^2}{{\mathcal H}} - 3 {\mathcal H}\biggr]  - \frac{{\mathcal H}^2 + 2 {\mathcal H}' + 4 \pi G {\varphi'}^2}{{\mathcal H}}\biggr\}\phi
\nonumber\\
&+& \frac{4 \pi G}{{\mathcal H}} \biggl[ ( 1 - w) \chi' \varphi' - a^2 \frac{\partial V}{\partial \varphi} ( 1 + w) \chi\biggr].
\label{QN11}
\end{eqnarray}
Equation (\ref{QN9}) will then be used into Eq. (\ref{QN11}) to eliminate $\phi$ and, therefore, the final form of Eq. (\ref{QN10}) is:
\begin{eqnarray}
&& u'' + {\mathcal H}(1 - 3 w) u' - w \nabla^2 u + \biggl\{ ( 1 - 3 w) {\mathcal H}' - \frac{4\pi G a^2(p_{pr} + \rho_{pr})}{{\mathcal H}^2} \biggl[ 4\pi G {\varphi'}^2 ( w -1) 
\nonumber\\
&& - \biggl( {\mathcal H}^2 ( 3 w + 1) + 2 {\mathcal H}'\biggr) \biggr] \biggr\}\, u
- \frac{4\pi G}{{\mathcal H}} \biggl\{\frac{\varphi'}{{\mathcal H}}\biggl[ 4\pi G {\varphi'}^2 (w-1) - \biggl( {\mathcal H}^2 ( 3 w + 1) + 2 {\mathcal H}'\biggr) \biggr] 
\nonumber\\
&& -  a^2 \frac{\partial V}{\partial \varphi} (w +1 ) \biggr\}\chi
 + \frac{4\pi G \varphi'}{{\mathcal H}} ( 1 - w) \chi'  =0.
\label{QN12}
\end{eqnarray}
The same strategy used to derive Eq. (\ref{QN12}) leads to the evolution equation of the second quasinormal mode associated with Eq. (\ref{chi}).
Here the problem is to trade the last three terms appearing at the right hand 
side of Eq. (\ref{chi}) for appropriate combinations of $\chi$, $u$ and $u'$. 
The relevant combination arising from the terms containing $\phi$, $\nabla^2 B$ 
and $\phi'$ can be transformed by using Eqs. (\ref{HG1}), (\ref{sep3}) and 
(\ref{QN9}). The final result is
\begin{eqnarray}
&& \chi'' + 2 {\mathcal H} \chi' - \nabla^2 \chi + 
\biggl\{ a^2 \frac{\partial^2 V}{\partial \varphi^2} + 4\pi G \biggl( \frac{\varphi'}{{\mathcal H}}\biggr) \biggl[ 4 a^2 \frac{\partial V}{\partial \varphi} + 
\biggl(\frac{\varphi'}{{\mathcal H}}\biggr) \biggl(2 ({\mathcal H}' + 2 {\mathcal H}^2) 
\nonumber\\
&& + 4\pi G a^2 (p_{pr} + \rho_{pr}) \biggl(\frac{w-1}{w}\biggr)\biggr)\biggr]\biggr\} \, \chi  - 4\pi G a^2 ( p_{pr} + \rho_{pr}) \biggl(\frac{\varphi'}{{\mathcal H}}\biggr) \biggl(\frac{w-1}{w}\biggr) u' 
\nonumber\\
&& + \frac{4\pi G a^2 ( \rho_{pr} + p_{pr})}{{\mathcal H}} \,\biggl\{ 2 \frac{\partial V}{\partial\varphi} a^2 + \biggl(\frac{\varphi'}{{\mathcal H}}\biggr)\biggl[ 2 ({\mathcal H}' + 2 {\mathcal H}^2) + 4\pi G a^2 (p_{pr} + \rho_{pr}) \biggl(\frac{w -1}{w}\biggr)\biggr] 
\nonumber\\
&& - {\mathcal H} \varphi' 
( 1 - 3 w) \frac{w -1}{w} \biggr\}\,u=0.
\label{QN13}
\end{eqnarray}
The derivation of Eqs. (\ref{QN12}) and (\ref{QN13}) breaks 
down when $c_{\mathrm{s}}^2 = \sqrt{w} = 0$ so that the latter case must be 
separately treated. The starting point of the derivation will be the same 
except that, since $\delta p_{pr}=0$, Eq. (\ref{th}) is even simpler. Repeating 
all the steps of the derivation we obtain, after some algebra, that the evolution 
equation for $u$ becomes:
\begin{eqnarray}
&& u''  + \biggl( {\mathcal H} - \frac{4 \pi G \rho a^2}{{\mathcal H}}\biggr) u'
+ \biggl[ {\mathcal H}' +\frac{ 4 \pi G a^2 \rho}{{\mathcal H}} \biggl( {\mathcal H} + \frac{{\mathcal H}'}{{\mathcal H}} \biggr) \biggr] u
\nonumber\\
&& + 4 \pi G \biggl(\frac{{\varphi'}}{{\mathcal H}}\biggr)\biggl[ 2 {\mathcal H} + \frac{{\mathcal H}'}{{\mathcal H}} + \frac{a^2}{{\varphi'}} \frac{\partial V}{\partial \varphi}\biggr] \chi - 4 \pi G \biggl(\frac{{\varphi'}}{{\mathcal H}}\biggr)\chi'=0.
\label{QN14}
\end{eqnarray}
Similarly, in the case $c_{\mathrm{s}}^2 = \sqrt{w} = 0$, the evolution equation 
for $\chi$ becomes, after some algebra:
\begin{eqnarray}
&& \chi'' + 2 {\mathcal H} \chi' - \nabla^2 \chi + \biggl\{ \frac{\partial^2 V}{\partial\varphi^2} a^2 + 8 \pi G \biggl(\frac{{\varphi'}}{{\mathcal H}}\biggr)\biggl[ 
2 \frac{\partial V}{\partial \varphi} a^2 + \biggl(\frac{{\varphi'}}{{\mathcal H}}\biggr)
\biggl( 2 + \frac{{\mathcal H}'}{{\mathcal H}^2} \biggr)\biggr]\biggr\} \chi  
\nonumber\\
&&+  \frac{8 \pi G a^2 \rho}{{\mathcal H}}
\biggl[ \frac{{\partial V}}{\partial \varphi} a^2 + 
\biggl(\frac{{\varphi'}}{{\mathcal H}}\biggr)\biggl( 2 + \frac{{\mathcal H}'}{{\mathcal H}^2} \biggr)\biggr] u=0
\label{QN15}
\end{eqnarray}

Going back to the main derivation and to  Eqs. (\ref{QN12}) and (\ref{QN13}), the expression for the evolution of $u$ can be modified by defining a new variable 
$v$ such that 
\begin{equation}
\phi = \frac{\varphi'}{2 {\mathcal H}} \chi + \frac{a \sqrt{p + \rho}}{2 {\mathcal H}} v, \qquad v = a \,\sqrt{p_{pr} + \rho_{pr}}\, u.
\label{QN16}
\end{equation}
In Eq. (\ref{QN16}) Planck units $ \overline{M}_{\mathrm{P}}^2 = 1$ have been set and, in these units, the equations 
for $v$ and $\chi$ become, respectively, 
\begin{eqnarray}
&& v'' + 2 {\mathcal H} v' - w \nabla^2 v + \biggl\{ \frac{3 (1 - w)}{2} {\mathcal H}' + \frac{3 ( 1 - w)}{4} {\mathcal H}^2 
\nonumber\\
&& - \frac{a^2 (p_{pr} + \rho_{pr})}{4 {\mathcal H}^2 }[ {\varphi'}^2 ( w -1) - 2 ( {\mathcal H}^2 (3 w +1) + 2 {\mathcal H}')]\biggr\} v 
\nonumber\\
 &&- \frac{a \sqrt{p_{pr} + \rho_{pr}}}{4 {\mathcal H}}\biggl\{ \biggl(\frac{\varphi'}{{\mathcal H}}\biggr) [{\varphi'}^2 ( w -1) - 2 ( {\mathcal H}^2 (3 w +1) + 2 {\mathcal H}')]
\nonumber\\
 && - 2 a^2 (w +1) \frac{\partial V}{\partial \varphi} \biggr\}\chi + \frac{a \sqrt{p + \rho}}{2 {\mathcal H}} \varphi' ( 1 - w) \chi' =0,
\label{QN17}
\end{eqnarray}
 and 
 \begin{eqnarray}
 && \chi'' + 2 {\mathcal H} \chi' - \nabla^2 \chi  + \biggl\{ a^2 \frac{\partial^2 V}{\partial \varphi^2} + \frac{\varphi'}{4 {\mathcal H}}\biggl[ 8 a^2 \frac{\partial V}{\partial \varphi} 
 \nonumber\\
&& + \frac{\varphi'}{{\mathcal H}}\biggl(4 ({\mathcal H}' + 2 {\mathcal H}^2) + a^2 ( p_{pr} + \rho_{pr}) \biggl(\frac{w -1}{w} \biggr) 
 \biggr)\biggr]\biggr\}\chi 
\nonumber\\ 
 && + \frac{a \sqrt{p_{pr} + \rho_{pr}}}{4 {\mathcal H}} \biggl\{ 3 \frac{(w -1)^2}{w}\varphi' {\mathcal H} + 4 a^2 \frac{\partial V}{\partial \varphi} + \frac{\varphi'}{{\mathcal H}} \biggl[ 4 ({\mathcal H}' + 2 {\mathcal H}^2)  
 \nonumber\\
&&   + a^2 (p_{pr} + \rho_{pr}) \biggl(\frac{w- 1}{w} \biggr)\biggr]\biggr\} v
 - \frac{a \sqrt{p_{pr} + \rho_{pr}}}{2 {\mathcal H}} \varphi' \biggl(\frac{w -1}{w} \biggr) v'=0.
 \label{QN18}
 \end{eqnarray}
 From Eqs. (\ref{QN17}) and (\ref{QN18}) the evolution equations for the canonical normal 
 modes
\begin{equation}
q_{v} = \frac{a\, v}{c_{\mathrm{s}}}, \qquad q_{\chi} = a \, \chi,
\label{QN19}
\end{equation}
leads to  Eqs. (\ref{NMM1}) and (\ref{NMM2}).  The variable $v$ is 
the closest analog, in the fluid case, to the inflaton fluctuation $\chi$. It is  
useful to recall, as a technical remark, that the intermediate steps involving the variable $v$ can be avoided. 
In the latter case $q_{v}$ is directly expressible in terms of the variable $u$ as $q_{v} = z_{pr}\, {\mathcal H} \, u$; with this 
strategy the algebraic derivation is, though, more tedious.

\end{appendix}


\newpage

\begin{thebibliography}{99}

\bibitem{wmap1} C.~L.~Bennett {\it et al.},  Astrophys.\ J.\ Suppl.\  {\bf 192}, 17 (2011);
N.~Jarosik {\it et al.},  Astrophys.\ J.\ Suppl.\  {\bf 192}, 14 (2011); 
 J.~L.~Weiland {\it et al.},  Astrophys.\ J.\ Suppl.\  {\bf 192}, 19 (2011).
 
 \bibitem{wmap2} D.~Larson {\it et al.}, Astrophys.\ J.\ Suppl.\  {\bf 192}, 16 (2011);
B.~Gold {\it et al.},  Astrophys.\ J.\ Suppl.\  {\bf 192}, 15 (2011);  
E.~Komatsu {\it et al.},   Astrophys.\ J.\ Suppl.\  {\bf 192}, 18 (2011).

\bibitem{one} P.~D.~B.~Collins and R.~F.~Langbein,ÊÊPhys.\ Rev.\ D {\bf 45}, 3429 (1992);  Phys.\ Rev.\ D {\bf 47}, 2302 (1993).

\bibitem{two} I.~Yu.~Sokolov, Class.\ Quant.\ Grav.\  {\bf 9}, L61 (1992).

\bibitem{three}  M.~Gasperini, M.~Giovannini, and G.~Veneziano,  Phys.\ Rev.\  {\bf D48}, 439-443 (1993).

\bibitem{four} K.~Bhattacharya, S.~Mohanty and A.~Nautiyal,  Phys.\ Rev.\ Lett.\  {\bf 97}, 251301 (2006); K.~Bhattacharya, S.~Mohanty and R.~Rangarajan, Phys.\ Rev.\ Lett.\  {\bf 96}, 121302 (2006).

\bibitem{five} W.~Zhao, D.~Baskaran and P.~Coles, Phys.\ Lett.\ B {\bf 680}, 411 (2009).

\bibitem{six}  M.~Giovannini,  Phys.\ Rev.\ D {\bf 83}, 023515 (2011).

\bibitem{seven} I.~Agullo and L.~Parker, ÊPhys.\ Rev.\ D {\bf 83}, 063526 (2011).

\bibitem{eight} R.~Lieu and T.~W.~B.~Kibble, arXiv:1110.1172 [astro-ph.CO].
  
\bibitem{nine}   S.~Kundu, JCAP {\bf 1202}, 005 (2012).

\bibitem{QO}  J.  Klauder and E. Sudarshan, {\it Fundamentals of quantum optics} (Benjamin, New York, 1968); 
 R. Loudon, {\it The quantum theory of light} (Clarendon Press, Oxford, 1983); 
  L. Mandel and E. Wolf, {\it Optical coherence and quantum optics}, (Cambridge University Press, Cambridge, 1995).

\bibitem{eleven} M.~Giovannini,  Class.\ Quant.\ Grav.\  {\bf 20}, 5455 (2003); Phys.\ Rev.\ D {\bf 73}, 083505 (2006).

\bibitem{bounce} M. Giovannini,  Class.\ Quant.\ Grav.\  {\bf 21}, 4209 (2004); Phys.\ Rev.\ D {\bf 70}, 103509 (2004); M.~Gasperini, M.~Giovannini and G.~Veneziano,  Phys.\ Lett.\ B {\bf 569}, 113 (2003); Nucl.\ Phys.\ B {\bf 694}, 206 (2004).

\bibitem{haw}  S.~W.~Hawking,  Phys.\ Lett.\  B {\bf 115}, 295 (1982); 
  A.~H.~Guth and S.~Y.~Pi,  Phys.\ Rev.\ Lett.\  {\bf 49}, 1110 (1982).

\bibitem{lukash} V.~N.~Lukash,  Sov.\ Phys.\ JETP {\bf 52}, 807 (1980) [Zh. Eksp. Teor. Fiz. {\bf 79}, 1601 (1980)].

\bibitem{lif} E.~M.~Lifshitz and I.~M.~Khalatnikov,
  Adv.\ Phys.\  {\bf 12}, 185 (1963); E.~M.~Lifshitz Zh. Eksp. Teor. Fiz. {\bf 16}, 587 (1946).

\bibitem{strokov}  V.~Strokov,  Astron.\ Rep.\  {\bf 51}, 431-434 (2007).

\bibitem{staro1} A.~A.~Starobinsky,  Phys.\ Lett.\ B {\bf 91}, 99 (1980);
  JETP Lett.\  {\bf 30}, 682 (1979) [Pisma Zh.\ Eksp.\ Teor.\ Fiz.\  {\bf 30}, 719 (1979)].
  
\bibitem{bard1} J. Bardeen, Phys. Rev. {\bf D22}, 1882 (1980).

\bibitem{KS}  H.~Kodama, M.~Sasaki,  Prog.\ Theor.\ Phys.\ Suppl.\  {\bf 78}, 1-166 (1984); 
M. Sasaki, Prog. Teor. Phys. {\bf 76}, 1036 (1986). 

\bibitem{chibisov} G.~V.~Chibisov, V.~F.~Mukhanov,  Mon.\ Not.\ Roy.\ Astron.\ Soc.\  {\bf 200}, 535 (1982); V.~F.~Mukhanov,  Sov.\ Phys.\ JETP {\bf 67}, 1297 (1988)  [Zh. Eksp. Teor. Fiz. {\bf 94}, 1 (1988)].

\bibitem{br1} R.~H.~Brandenberger, R.~Kahn and W.~H.~Press,
  Phys.\ Rev.\ D {\bf 28}, 1809 (1983);  R.~H.~Brandenberger and R.~Kahn,  Phys.\ Rev.\ D {\bf 29}, 2172 (1984).

\bibitem{bard2} J. Bardeen, P. Steinhardt, and M. Turner, Phys. Rev. {\bf D28}, 679 (1983); 
J.~A.~Frieman and M.~S.~Turner, Phys.\ Rev.\ D {\bf 30}, 265 (1984).

\bibitem{hoff1} J~-c.~Hwang, Astrophys. J.  {\bf 375}, 443 (1990);
 J.~-c.~Hwang,  Class.\ Quant.\ Grav.\  {\bf 11}, 2305 (1994); J.~-c.~Hwang and H.~Noh,  Phys.\ Lett.\ B {\bf 495}, 277 (2000);
J.~c.~Hwang and H.~Noh,  Class.\ Quant.\ Grav.\  {\bf 19}, 527 (2002).

\bibitem{trans} J.~-c.~Hwang and E.~T.~Vishniac, Astrophys.\ J.\  {\bf 382}, 363 (1991);  N.~Deruelle and V.~F.~Mukhanov,
 Phys.\ Rev.\ D {\bf 52}, 5549 (1995); E.~J.~Copeland and D.~Wands,
  JCAP {\bf 0706}, 014 (2007).

\bibitem{LLS}  S.~M.~Leach, M.~Sasaki, D.~Wands and A.~R. Liddle,
  Phys.\ Rev.\ D {\bf 64}, 023512 (2001); S.~M.~Leach and A.~R.~Liddle,
  Phys.\ Rev.\ D {\bf 63}, 043508 (2001).

\bibitem{grad1} K. Tomita, Prog. Theor. Phys.  {\bf 67}, 1076 (1982);
K. Tomita, Phys. Rev. D {\bf 48}, 5634 (1993).

\bibitem{grad2} N. Deruelle and D. Goldwirth, Phys. Rev. D {\bf 51}, 1563 (1995);
N. Deruelle and K. Tomita, Phys. Rev. D {\bf 50}, 7216 (1994);
G. Comer, N. Deruelle, D. Langlois, and J. Parry, Phys. Rev. D {\bf 49}, 2759 (1994).

\bibitem{grad3}  M.~Giovannini, JCAP {\bf 0509}, 009 (2005); M.~Giovannini and Z.~Rezaei,
  Phys.\ Rev.\ D {\bf 83}, 083519 (2011); Class.\ Quant.\ Grav.\  {\bf 29}, 035001 (2012).



\end{thebibliography}
\end{document}